%% file: main.tex
\documentclass[sigconf]{acmart}
\AtBeginDocument{%
  \providecommand\BibTeX{{%
    \normalfont B\kern-0.5em{\scshape i\kern-0.25em b}\kern-0.8em\TeX}}}

\setcopyright{acmcopyright}
\copyrightyear{2018}
\acmYear{2018}
\acmDOI{XXXXXXX.XXXXXXX}

\acmConference[Conference acronym 'XX]{Make sure to enter the correct
  conference title from your rights confirmation emai}{June 03--05,
  2018}{Woodstock, NY}
\acmPrice{15.00}
\acmISBN{978-1-4503-XXXX-X/18/06}

\usepackage{color}

\definecolor{blush}{rgb}{0.87, 0.36, 0.51}

\usepackage[linesnumbered,ruled]{algorithm2e}
\usepackage{wrapfig}

\usepackage{multirow}

\usepackage{xcolor}

\definecolor{LightBlue}{HTML}{6EA6D5}
\definecolor{DarkBlue}{HTML}{0000A0}
\definecolor{MediumBlue}{HTML}{000000} 
\definecolor{MyColor}{HTML}{000000} 
\newcommand{\hl}[1]{\textcolor{MyColor}{#1}}

\newcommand{\mynoindent}{\vspace{0.3em} \noindent}
\newcommand{\boldstart}[1]{\mynoindent \textbf{#1}}

\definecolor{avinashcolor}{rgb}{0.37, 0.29, 0.95}

\newcommand{\frmname}{{\sf Reshape}\xspace}

\mathchardef\mhyphen="2D

\usepackage[labelfont=bf]{caption}
\usepackage{subcaption}

\usepackage{amsmath} 

\usepackage{tabularx}

\begin{document}

\title{ Reshape: Adaptive Result-aware Skew Handling for Exploratory
Analysis on Big Data}
\author{Avinash Kumar, Sadeem Alsudais, Shengquan Ni, Zuozhi Wang, Yicong Huang, Chen Li}
\affiliation{%
  \institution{Department of Computer Science, UC Irvine, CA 92697, USA}
}
\email{{avinask1, salsudai, shengqun, zuozhiw, yicongh1, chenli}@ics.uci.edu}

\input{abstract}



\maketitle

\input{sec1}
\input{sec2}
\input{sec3}
\input{sec4}
\input{sec5}
\input{sec6}
\input{sec7}
\input{sec8}


\bibliographystyle{ACM-Reference-Format}
\bibliography{references,edbt2023}










\end{document}

%% file: abstract.tex
\begin{abstract}

The process of data analysis, especially in GUI-based analytics systems, is highly exploratory. The user iteratively refines a workflow multiple times before arriving at the final workflow. In such an exploratory setting, it is valuable to the user if the initial results of the workflow are representative of the final answers so that the user can refine the workflow without waiting for the completion of its execution. Partitioning skew may lead to the production of misleading initial results during the execution. In this paper, we explore skew and its mitigation strategies from the perspective of the results shown to the user. We present a novel framework called \frmname that can adaptively handle partitioning skew in pipelined execution. \frmname employs a two-phase approach that transfers load in a fine-tuned manner to mitigate skew iteratively during execution, thus enabling it to handle changes in input-data distribution. \frmname has the ability to adaptively adjust skew-handling parameters, which reduces the technical burden on the users. \frmname supports  a variety of operators such as {\sf HashJoin}, {\sf Group-by}, and {\sf Sort}. We implemented \frmname on top of two big data engines, namely Amber and Flink, to demonstrate its generality and efficiency, and report an experimental evaluation using real and synthetic datasets.

\end{abstract}


%% file: sec1.tex
\section{Introduction}
\label{sec:introduction}

As information volumes in many applications become large, data analytics is becoming increasingly important. Data processing frameworks such as Hadoop~\cite{misc/hadoopmapreduce}, Spark~\cite{misc/spark}, and Flink~\cite{misc/flink} provide programming interfaces that are used by developers to code their data processing needs. GUI-based workflow systems such as Alteryx~\cite{alteryx}, RapidMiner~\cite{rapidMiner}, Knime~\cite{knime}, Einblick~\cite{Einblick}, and Texera~\cite{journals/pvldb/WangKNL20} provide a GUI interface where the users can drag-and-drop operators and create a workflow as a directed acyclic graph (DAG). Once the data processing job is created, it is submitted to an engine that executes the job. 

\begin{figure}[htbp]
	\vspace{-0.12in}
	\includegraphics[width=3.2in]{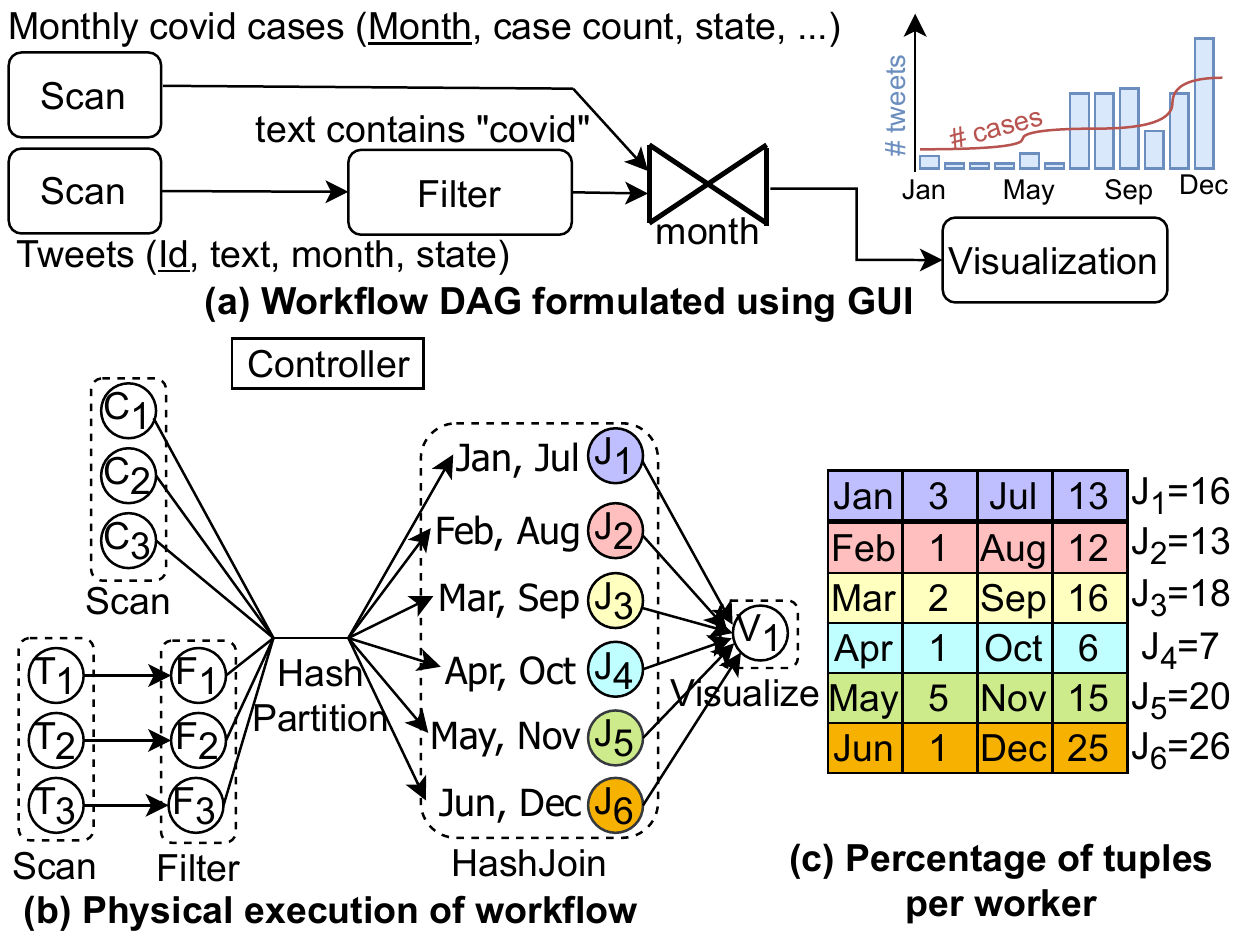}
	\vspace{-0.12in}
	\caption{\label{fig:tweets-workflow}
		\textbf{Partitioning skew in a data science project of Covid tweet analysis.}
	}
\end{figure}

The process of data analysis, especially in GUI-based analytics systems, has two important characteristics. \textbf{1) Highly exploratory:} The process of building a workflow can be very exploratory and iterative~\cite{journals/interactions/FisherDCD12, conf/sigmod/XuKAR22,conf/sigmod/VartakSLVHMZ16}. Often the user constructs an initial workflow and executes it to observe a few results. If they are not desirable, she terminates the current execution and revises the workflow. The user iteratively refines the workflow until finishing a final workflow to compute the results. As an example, Figure~\ref{fig:tweets-workflow} shows a workflow at an intermediate step during the task of covid data analysis. It examines the relationship between the number of tweets containing the keyword {\tt covid} and the number of Covid cases in 2020. The monthly details about the Covid cases are joined with tweets filtered on the {\tt covid} keyword on the month column. The result is plotted as a visualization operator that shows a bar chart about the total count of tweets about Covid per month and a line chart about the total Covid cases per month. The analyst may observe the visualization and choose to continue refining the workflow to do analysis for specific US states. In the Texera system we are developing, we observe that the users refined a workflow about $80$ times on an average before reaching the final version. \textbf{2) Suitable for non-technical users:} GUI-based workflow execution systems significantly lower the technical learning curve for its users, thus enabling non-IT domain experts to do data science projects without writing code. Such systems also try to minimize the requirements on users to know the technical details of workflow execution, so that the user can focus solely on the analytics task at hand.


In exploratory data analytics, it is vital for a user to see results quickly to allow her to identify problems in the analysis early and take corrective actions without waiting for the entire workflow to finish executing. {\em Pipelined execution}~\cite{journals/csur/BenoitCRS13} is a popular workflow execution model that can produce results quickly. In pipelined execution, an operator does not wait for its entire input data to be produced before processing the input and sending results to its downstream operators. For example, when the workflow in Figure~\ref{fig:tweets-workflow} is executed using pipelined model, the {\sf HashJoin} operator starts executing and producing results as soon as the {\sf Filter} operator outputs initial results. The user can notice the initial results and make any changes, if needed.  Pipelined execution is adopted by data-processing engines such as Flink~\cite{misc/flink}, Samza~\cite{Samza:website}, and Storm~\cite{apacheStorm}.

As data volumes in these systems increase, it is indispensable to do parallel processing, in which data is partitioned and processed by multiple computation units in parallel. Data partitioning, either using hash partitioning or range partitioning, often results in skew.  As an example, the {\sf HashJoin} operator in Figure~\ref{fig:tweets-workflow} receives hash partitioned inputs from the two upstream operators.  Although the hash function allots the same number of months to each join worker, load imbalance still exists because of different numbers of tweets for those months. It is well known that partitioning skew adversely affects the efficiency of engines as it increases the processing time and reduces the throughput~\cite{conf/vldb/DeWittNSS92,conf/sigmod/KwonBHR12}.

The problem of partitioning skew has been extensively studied in the literature, mainly from the perspective of increasing the end-to-end performance. However, there is little research on the following important problem:

\begin{quote}
{\em In exploratory data analytics, how to consider the results shown to the user when mitigating skew?}
\end{quote} 

In exploratory data analysis, it is valuable to the analyst if the initial results are representative of the final results because they allow her to identify issues early and make necessary changes. Partitioning skew may lead to the production of misleading results during the execution. Let us consider the production rate of October and December tuples from the {\sf HashJoin} operator in the running example. Assume that the {\sf HashJoin} operator is the bottleneck of the execution, and its workers receive input at an equal or higher rate than they can process. Although there are more December tuples than October, their production rates are similar because the total amounts of data received by $J_4$ and $J_6$ are different (details in Section~\ref{sec:load-transfer-mechanism}). Thus, the bar chart shows similar heights for October and December bars till $J_4$ completes processing, whereas the December bar is almost four times taller than the October bar in the final result. In this paper, we explore partitioning skew mitigation in the setting of exploratory data analysis and analyze the effect of mitigation strategies on the results shown to the user.

A common solution to handle partitioning skew at an operator is blocking the partitioning of its input data till the entire input data is produced by its upstream operator~\cite{SparkAQE:website,conf/vldb/DeWittNSS92,conf/icde/VitorovicE016,journals/tpds/ChenYX15,conf/sigmod/AbdelhamidMDA20} and then sampling the input data to create an optimal partitioning function. For example, in Figure~\ref{fig:tweets-workflow}, the {\sf HashJoin} operator waits for the {\sf Filter} operator to completely finish. Then, the output of the {\sf Filter} operator is sampled to create an optimal partitioning function to send data to the {\sf HashJoin} operator. Such blocking is not allowed in pipelined execution which makes these solutions infeasible in pipelined execution setting. Even temporarily blocking the partitioning till a small percentage of input (e.g., 1\%~\cite{conf/icde/RodigerIK016}) is collected for sampling can result in a long delay if there is an expensive upstream operator.


A different solution applicable to the pipelined execution setting is to detect the overloaded workers of an operator at runtime and transfer the processing of a few keys of the overloaded worker to a more available worker. For example, $J_6$ is detected to be overloaded at runtime and the processing of June tuples is transferred to $J_4$. However, this transfer has little effect on the results shown to the user (details in Section~\ref{sec:load-transfer-mechanism}). In order to show representative initial results, the data of December has to be split between $J_4$ and $J_6$. Thus, these two approaches of transferring load from $J_6$ to $J_4$ have different impacts on the initial results shown to the user.




In this paper, we analyze the effect of different skew mitigation strategies on the results shown to the user and present a novel skew handling framework called \frmname that adaptively handles skew in a pipelined execution setting. Reshape monitors the workload metrics of the workers and adapts the partitioning logic to transfer load whenever it observes a skew.  These modifications can be done multiple times as the input distribution changes~\cite{conf/sigir/BeitzelJCGF04,conf/wsdm/KulkarniTSD11} or if earlier modifications did not fully mitigate the skew. The command to adapt the partitioning logic is sent from the controller to the workers using low latency control messages that are supported in various engines such as Flink, Chi~\cite{journals/pvldb/MaiZPXSVCKMKDR18}, and Amber~\cite{journals/pvldb/KumarWNL20}. 

We make the following contributions. (1) Analysis of the impact of mitigation on the shown results: We present different approaches of skew mitigation and analyze their impact on the results shown to the user. (Section~\ref{sec:load-transfer-mechanism}). (2) Automatic adjustment of the skew detection threshold: We present a way to dynamically adjust the skew detection threshold to reduce the number of iterations of mitigation to minimize the technical burden on the user
(Section~\ref{sec:adaptive-handling}). (3) Applicability to multiple operators: Since a data analysis workflow can contain many operators that are susceptible to partitioning skew, we generalize \frmname to multiple operators such as {\sf HashJoin}, {\sf Group-by}, and {\sf Sort}, and discuss challenges related to state migration (Section~\ref{sec:other-operators}). (4) Generalization to broader settings: We consider settings such as high state-migration time and multiple helper workers for an overloaded worker and discuss how \frmname can be extended in these settings (Section~\ref{sec:broader-settings}).  (5) Experimental evaluation: We present the implementation of \frmname on top of two big-data engines, namely Amber and Flink, to show the generality of this approach. We report an experimental evaluation using real and synthetic datasets on large computing clusters (Section~\ref{sec:experiments}).

\subsection{Related work}
\label{ssec:related-work}

There have been extensive studies about skew handling in two major execution paradigms in big data engines -- batch execution and pipelined execution. Batch-execution systems such as MapReduce~\cite{misc/hadoopmapreduce} and Spark~\cite{misc/spark} materialize complete input data before partitioning it across workers. Pipelined-execution systems such as Flink~\cite{misc/flink}, Storm~\cite{apacheStorm} and Amber~\cite{journals/pvldb/KumarWNL20} send input tuples to a receiving worker immediately after they are available. The complete input is not known to the operator in pipelined execution, which makes skew handling more challenging.

\boldstart{Skew handling in batch execution.}  A static technique is to sample and obtain the distribution of complete input data and use it to partition data in a way that avoids skew~\cite{SparkAQE:website, conf/vldb/DeWittNSS92, conf/icde/VitorovicE016,journals/tpds/ChenYX15,conf/sigmod/AbdelhamidMDA20}. Adaptive skew-handling techniques adapt their decisions to changing runtime conditions and mitigate skew in multiple iterations. For instance, SkewTune~\cite{conf/sigmod/KwonBHR12} and Hurricane~\cite{conf/eurosys/BindschaedlerMS18} handle skew adaptively. That is, upon detecting skew, SkewTune stops the executing workers, re-partitions the materialized input, and starts new workers to process the partitions. Hurricane clones overburdened workers and uses a special storage that allows fine-grained data access to the original and cloned workers in parallel. Hurricane can split the processing of a key over multiple workers and thus has a fine load-transfer granularity. SkewTune cannot split the processing of a key.




\boldstart{Static skew handling in pipelined execution.} Flow-Join~\cite{conf/icde/RodigerIK016} avoids skew in a {\sf HashJoin} operator. It samples the first 1\% of input data of the operator to decide the overloaded keys and does a broadcast join for the overloaded keys. Since it makes the decision based on an initial portion of the input, it cannot handle skew if the input distribution changes multiple times during the execution. Partial key grouping (PKG)~\cite{conf/icde/NasirMGKS15,conf/icde/NasirMKS16} uses multiple pre-defined partitioning functions. It results in multiple candidate workers sharing the processing of the same key. Since the partitioning logic is static, a worker may process multiple skewed keys, which makes it more burdened than other workers. PKG cannot be used to handle skew in operators such as {\sf Sort} and {\sf Median}.

\boldstart{Adaptive skew handling in pipelined execution.} Flux~\cite{conf/icde/ShahHCF03} divides the input into many pre-defined mini-partitions that can be transferred between workers to mitigate skew. Thus, the load-transfer granularity is fixed and pre-determined. Also, it cannot split the load of a single overloaded key to multiple workers. Another adaptive technique minimizes the input load on workers that compute theta joins by dynamically changing the replication factor for data partitions~\cite{journals/pvldb/ElseidyEVK14}. This approach uses random partitioning schemes such as round-robin and hence is not prone to partitioning skew. \frmname handles skew adaptively over multiple iterations. It determines the keys to be transferred dynamically and allows an overloaded key to be split over multiple workers for mitigation.


%% file: sec2.tex
\section{\frmname: Overview}
\label{sec:overview}

\begin{figure}[htbp]
    \vspace{-0.12in}
	\includegraphics[width=3.2in]{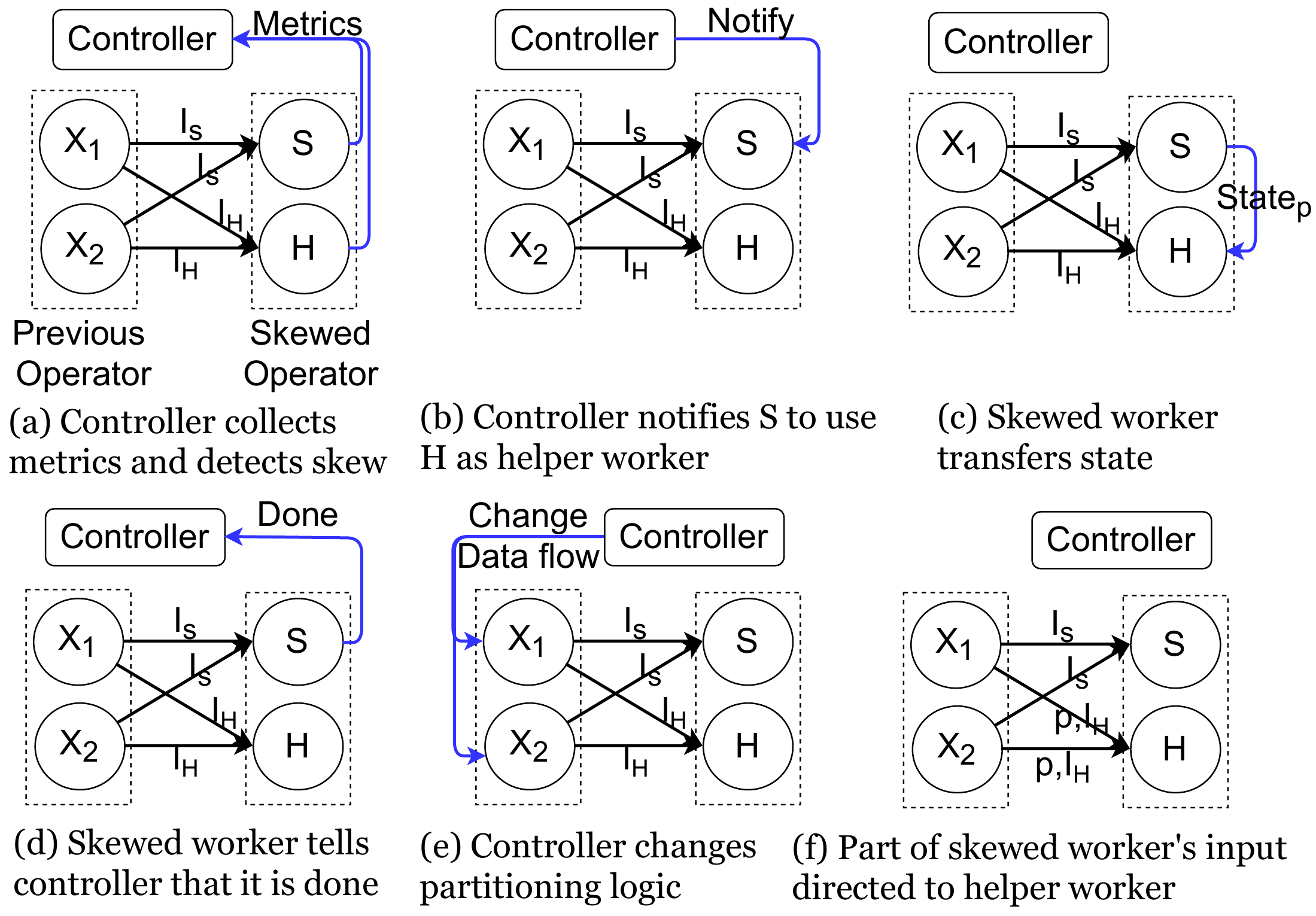} 
	\vspace{-0.12in}
	\caption{\label{fig:overall-approach}
		\textbf{Steps of skew-handling in \frmname. Skew detected in (a) and mitigated in (b)-(f).}
	}
    \vspace{-0.2in}
\end{figure}

We use Figure~\ref{fig:overall-approach} to give an overview of \frmname.


\subsection{Skew detection} 
\label{ssec:skew-detection}
During the execution of an operator, the controller periodically collects workload metrics from the workers of the operator to detect skew (Figure~\ref{fig:overall-approach}(a)). There are different metrics that can represent the workload on a worker such as CPU usage, memory usage and unprocessed data queue size~\cite{conf/eurosys/BindschaedlerMS18, conf/sigmod/FernandezMKP13}.  Skew handling in \frmname is independent of the choice of workload metric, and we choose unprocessed queue size as a metric in this paper. We choose this metric because the results seen by the analyst depend on the future results produced by a worker, which in turn depend on the content of its unprocessed data queue. We refer to a computationally overburdened worker as a {\em skewed worker} and workers that share the load as its {\em helper workers}. 

\boldstart{Skew test.} Given two workers of the same operator, say $C$ and $L$, the controller performs a {\em skew test} to determine whether $C$ is a helper candidate for $L$. The skew test uses the following inequalities to check if $L$ is computationally burdened and the workload gap between $L$ and $C$ is big enough:
\begin{equation}\label{eq:greater-than-eta}
    \phi^{}_L \geq \eta,
\end{equation}
\vspace{-0.22in}
\begin{equation}\label{eq:tau-diff}
    \phi^{}_L - \phi^{}_C \geq \tau,
\end{equation}
where $\eta$ and $\tau$ are threshold parameters and $\phi^{}_w$ is the workload on a worker $w$. 

\boldstart{Helper workers selection.} The skew tests may yield multiple helper candidates for $L$. For simplicity, we assume till Section~\ref{sec:other-operators} that one helper worker is assigned per skewed worker. In Section~\ref{sec:broader-settings}, we generalize the discussions by considering multiple helpers per skewed worker. The controller chooses the helper candidate with the lowest workload that has not been assigned to any other overloaded worker as the helper of $L$. In our discussions forward, we use $S$ and $H$ to refer to a skewed worker and its chosen helper worker respectively.

\subsection{Skew mitigation} 
\label{ssec:skew-mitigation}
Suppose the skewed worker $S$ and its helper $H$ have been handling input partitions $I^{}_S$ and $I^{}_H$, respectively (Figure~\ref{fig:overall-approach}(a)). \frmname transfers a fraction of the future input of $S$ to $H$ to reduce the load on $S$. Here future input refers to the data input that is supposed to be received by a worker but has not yet been sent by the previous operator. The controller notifies $S$ about the part {\em p} of partition $I^{}_S$ that will be shared with $H$ to reduce the load of $S$ (Figure~\ref{fig:overall-approach}(b)). The downstream results shown to the user have a role to play in deciding {\em p}, which will be discussed in Section~\ref{sec:load-transfer-mechanism}. Worker $S$ sends to $H$ its state information $State_p$ corresponding to the partition {\em p}  (Figure~\ref{fig:overall-approach}(c)). Details about state-migration strategies are in Section~\ref{sec:other-operators}. We assume the state migration time to be small till Section~\ref{sec:other-operators}. In Section~\ref{sec:broader-settings}, we consider the general case where the state-migration time can be significant. Worker $H$ saves the state information and sends an {\em ack} message to $S$, which then notifies the controller (Figure~\ref{fig:overall-approach}(d)). The controller changes the partitioning logic at the previous operator (Figure~\ref{fig:overall-approach}(e,f)).

\boldstart{Fault Tolerance.} The \frmname framework supports the fault tolerance mechanism of the Flink engine~\cite{journals/corr/CarboneFEHT15} that checkpoints the states of the workers periodically. During checkpointing, a checkpoint marker is propagated downstream from the source operators. When an operator receives the marker from all its upstream operators, it takes a checkpoint which saves the current states of the workers of the operator. Every checkpoint has the information about the current partitioning logic at the workers. If checkpointing occurs during state migration, then the skewed worker additionally forwards the checkpoint marker to each of its helper workers. A helper worker needs to wait for the checkpoint marker from its corresponding skewed worker. Since the skewed workers and the helper workers are two disjoint sets of workers, there is no cyclic dependency in marker propagation and the checkpointing process successfully terminates. During recovery, the workers restore their states from the most recent checkpoint and then continue the execution.

%% file: sec3.tex
\section{Result-aware Load transfer}
\label{sec:load-transfer-mechanism}

After helper workers are selected for the skewed workers, the load needs to be transferred from the skewed workers to the corresponding helper workers. In Section~\ref{ssec:two-load-transfer-approaches}, we consider the different approaches of load transfer between workers and analyze their impact on the results shown to the user. Unlike other skew handling approaches that focus on evenly dividing the future incoming load among the workers, \frmname has an extra phase of load transfer at the beginning that removes the existing load imbalance between the workers. In Section~\ref{ssec:two-phases}, we discuss these two phases of load transfer and the significance of the first phase. 



\subsection{Mitigation impact on user results}
\label{ssec:two-load-transfer-approaches}

There are broadly two approaches to transfer the load from a skewed worker to its helper worker. We use the probe input of the {\sf HashJoin} operator in Figure~\ref{fig:tweets-workflow} (from the {\sf Filter} operator) as an example to explain the concepts in this section. It is assumed that the build phase of the join has finished. Suppose \frmname detects $J_6$ and $J_5$ as the skewed workers in the running example and $J_4$ and $J_2$ are their corresponding helpers, respectively.  The load-transfer approaches are implemented by changing the partitioning logic at the {\sf Filter} operator and affects the future tuples going into the {\sf HashJoin} operator.

{\em 1. Split by keys (SBK).} In this approach, the keys in the partition of the skewed worker are split into two disjoint sets, say $p_1$ and $p_2$. The future tuples belonging to $p_2$ are redirected to the helper worker, while tuples belonging to $p_1$ continue to be sent to the skewed worker. For example, the partition of the skewed worker $J_6$ is divided into $p_1$ = \{December\} and $p_2$ = \{June\}, and the future June tuples are sent to $J_4$, while December tuples continue to go to $J_6$.


{\em 2. Split by records (SBR).} In this approach, the records of the keys in the partition of the skewed worker are split between the skewed and the helper worker. The ratio of the split decides the amount of load transferred to the helper worker. For example, if the {\sf Filter} operator needs to redirect $\frac{9}{26}$ of the input $J_6$ to $J_4$, then it redirects $9$ tuples out of every $26$ tuples in $J_6$'s partition to $J_4$.

\boldstart{Impact of the two approaches on user results.} The two load transfer approaches have their own advantages and limitations. For example, {\sf SBK} incurs an extra overhead compared to {\sf SBR} because {\sf SBK} requires the workers to store the distribution of workload per key. On the other hand, {\sf SBR} may require transfer of a larger state size compared to {\sf SBK}, if all the keys of a skewed worker are shared with the helper. There are existing works in literature that address these concerns~\cite{conf/icdt/MetwallyAA05, conf/icde/RodigerIK016, conf/bigdataconf/YanXM13, conf/icde/GuflerARK12, conf/sigmod/MonteZRM20, journals/pvldb/HoffmannLMKLR19}. In the remainder of this subsection, we compare these two approaches from the perspective of their effects on the results shown to the user.

\begin{figure}[htbp]
    \vspace{-0.1in}
	\includegraphics[width=2.8in]{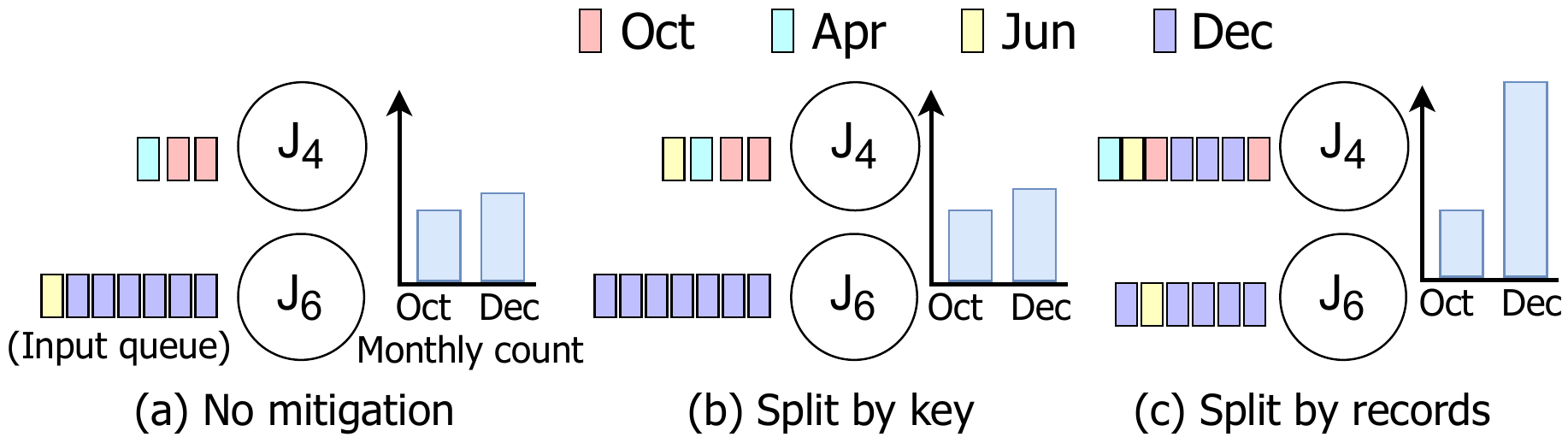}
	\vspace{-0.12in}
	\caption{\label{fig:split-by-record-better}
		\textbf{SBR splits December tuples on both workers and shows representative bar charts.}
	}
\end{figure}

\boldstart{a) Representative initial results.} As discussed before, it is valuable to the user if the initial results are representative of the final results. Partitioning skew may lead to the production of misleading results during the execution as shown next. Let us consider the bar chart visualization for October and December in the running example. The total count of December tweets, according to Figure~\ref{fig:tweets-workflow}(c), is about four times that of October tweets, i.e., the December bar is about four times longer than the October bar in the final visualization. Assume that the join operator is the bottleneck of the execution, and its workers receive input at an equal or higher rate than what they can process. Also assume that the processing speeds of the workers of {\sf HashJoin} are the same, say $t$ per second. $J_4$ produces $\frac{6}{7}*t$ October tuples and $J_6$ produces $\frac{25}{26}*t$ December tuples per second in the unmitigated case (Figure~\ref{fig:split-by-record-better}(a)). The rate of production of October and December tuples are similar because the total amount of data received by $J_4$ and $J_6$ are different. The bar chart shows similar heights for October and December bars in the unmitigated case till $J_4$ completes its processing.

When {\sf SBK} is used to mitigate the skew, the processing of June tuples is transferred to $J_4$ (Figure~\ref{fig:split-by-record-better}(b)). However, this transfer has little effect on the results shown to the user. The production rates of October and December after the transfer are $\frac{6}{8}*t$ and $t$ respectively.  That is, the heights of the December and October bars are still about the same, which is not representative of the final results.

{\sf SBR} has more flexibility for transferring load than {\sf SBK} because {\sf SBR} can split the tuples of a key over multiple workers. It leads to more representative initial results than {\sf SBK} as shown next. The processing of December and June tuples can be split between $J_6$ and $J_4$. For simplicity of calculation, we assume that only December tuples are shared with $J_4$. Since December tuples are now processed by two workers instead of one, the speed of production of these tuples increases. In order to make the future workloads of $J_4$ and $J_6$ similar, {\sf SBR} redirects $\frac{9}{26}$ of the input of $J_6$ to $J_4$, which increases the total percentage load on $J_4$ to $16$ and decreases that on $J_6$ to $17$. This is implemented by redirecting $9$ December tuples out of every $26$ tuples in $J_6$'s partition to $J_4$. The production rates of October tuples after the transfer is $\frac{6}{16}*t$. The December tuples are produced by $J_4$ and $J_6$. The production rate of December by $J_4$ is $\frac{9}{16}*t$ and by $J_6$ is $\frac{16}{17}*t$, which results in a total of approximately $\frac{24}{16}*t$. Thus, using {\sf SBR} leads to a more representative production ratio of December to October tuples of about $24{\,:\,}6$, which is similar to the actual ratio of $25{\,:\,}6$.

\begin{figure}[htbp]
    \vspace{-0.1in}
	\includegraphics[width=2.8in]{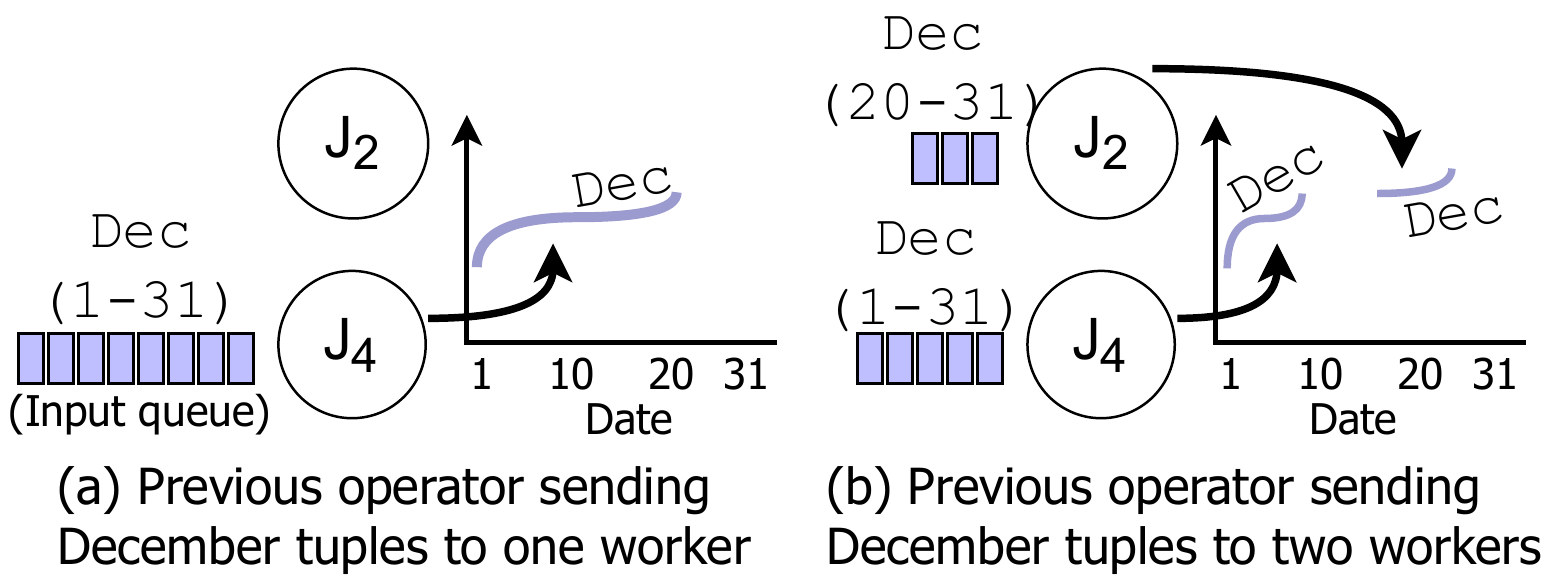}
	\vspace{-0.12in}
	\caption{\label{fig:split-by-key-better}
		\textbf{Processing a key at multiple workers by {\sf SBR} leads to a broken line chart. Only December tuples have been shown for simplicity.}
	}
\end{figure}

\boldstart{b) Preserving order of tuples.} If the tuples of a key being input into an operator are in a particular order and they need to be processed in that order, then {\sf SBK} is the suitable approach because it enforces a processing order by restricting the processing of the tuples of a key to a single worker at a time. If the processing of a key needs to be transferred to another worker, the migration can be synchronized using techniques such as pause and resume~\cite{conf/sigmod/ArmbrustDTYZX0S18,journals/pvldb/CarboneEFHRT17,conf/icde/ShahHCF03} or markers~\cite{journals/pvldb/ElseidyEVK14} (details in Section~\ref{sec:other-operators}) so that the tuples are processed in order. In contrast, {\sf SBR} distributes the tuples of a key over multiple workers to be processed simultaneously, which may cause them to be processed out of order. Consider the following example where an out-of-order processing of the tuples of a key is not desirable. Let us slightly modify the visualization operator in the running example to plot a line chart that shows daily count of covid related tweets. The daily count for each month is plotted as a separate line in the line chart. Figure~\ref{fig:split-by-key-better} shows the plot for  December in the line chart. Applications may want to show such plots as a continuous line with no breaks, starting from day 1 and extending towards increasing dates as execution progresses, for user experience purposes~\cite{Cloudberry}. In order to achieve this, the tuples of a month input into the {\sf HashJoin} operator are sorted in the increasing order of date. It is expected that {\sf HashJoin} produces tuples sorted by date, which can be consumed by the visualization operator to create a continuous plot.

{\sf SBK} assures that the December key is processed by only one worker at a time. Thus, it preserves the order of December tuples in the output sent to the visualization operator (Figure~\ref{fig:split-by-key-better}(a)). When {\sf SBR} is used, the December tuples are split between $J_4$ and $J_2$. In the example shown in Figure~\ref{fig:split-by-key-better}(b), the {\sf Filter} operator starts partitioning December tuples by {\sf SBR} when the tuples around the $20^{th}$ of December are being produced by the {\sf Filter} operator. Consequently, $J_2$ starts receiving the tuples from the date of the $20^{th}$ December and above. As $J_2$ and $J_4$ concurrently process data, the visualization operator receives the tuples out of order, resulting in broken line chart plots as shown in the figure.

In conclusion, {\sf SBR} allows more flexibility and enables the production of representative initial results than {\sf SBK}, but {\sf SBR} does not preserve the order of tuples. Thus, {\sf SBR} can be chosen unless there exists a downstream operator that imposes some requirement over the input order of the tuples. Such operators can be found at the workflow compilation stage. The operators before such an operator in the workflow can adopt {\sf SBK}.


\subsection{Extra phase in load transfer}
\label{ssec:two-phases}

The goal of skew mitigation is to use one of the two  approaches to transfer the load from the skewed worker to the helper worker in such a way that both workers have a similar workload for the rest of the execution. The skew handling works in literature usually have a single phase of load transfer that focuses on splitting the incoming input such that the workers receive similar load in future. \frmname has an extra phase of load transfer at the beginning that removes the existing load imbalance between the workers. We first give an overview of the two phases in \frmname, and explain the significance of the first phase.

\begin{figure}[htbp]
    \vspace{-0.1in}
	\includegraphics[width=3.3in]{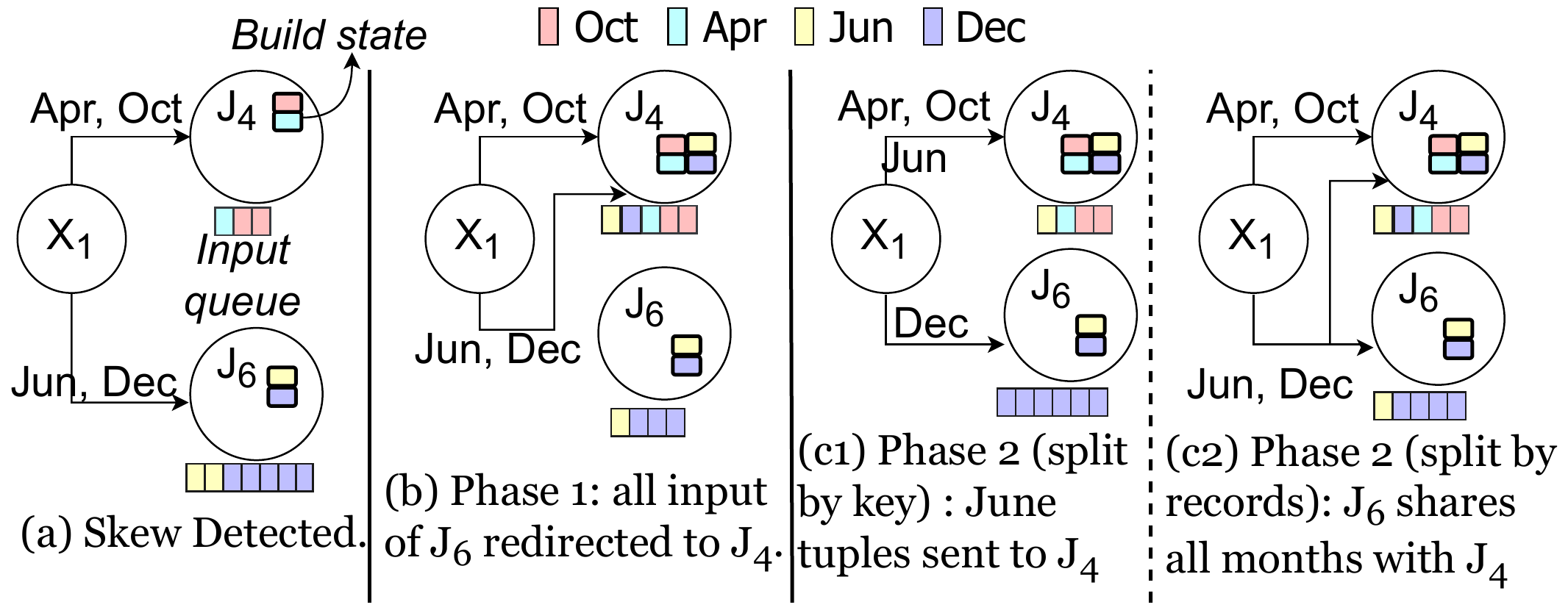}
	\vspace{-0.12in}
	\caption{\label{fig:phases-of-mitigation}
		\textbf{An implementation of the two phases using the ``SBK'' and ``SBR'' approach. $X_1$ is a previous operator worker.}
	}
\end{figure}

\boldstart{First Phase.} After the detection of skew (Figure~\ref{fig:phases-of-mitigation}(a)), the controller starts the first phase of load transfer. The first phase lets the helper ``catch up'' quickly with the skewed worker. One implementation of the partitioning logic in the first phase at the {\sf Filter} operator is that it sends all future tuples of $J_6$ to $J_4$ (Figure~\ref{fig:phases-of-mitigation}(b)). Note that $J_6$ will continue to process the data in its queue.  An alternative implementation is to send only a portion of $J_6$'s partition, such as the December data, to $J_4$. This alternative reduces the amount of state transfer, but it will take longer time for $J_4$ to catch up with $J_6$.

\boldstart{Second Phase.} Once the queue sizes of the two workers become similar, the controller starts the second phase. Its goal is to modify the partitioning logic at the {\sf Filter} operator to redirect part of the future input of $J_6$ in such a way that both the workers receive a comparable workload. In order to do this, first the incoming workload of the workers needs to be estimated. A sample of workloads needs to be collected to estimate the future workload of the workers~\cite{conf/cloud/RamakrishnanSU12, journals/tpds/ChenYX15, conf/bigdataconf/YanXM13, conf/icde/GuflerARK12} using a prediction function~$\psi$. \frmname can use the sample from the recent history collected during the current execution~\cite{conf/IEEEcloud/Kim0QH16, conf/cloud/ShenSGW11}.  If historical data is available, it can complement the recent data and improve the prediction accuracy~\cite{conf/isorc/GarraghanOTX15, conf/icde/PopescuEBBA12}.

To simplify the discussion of the second phase, we make the following assumptions: 
\begin{itemize}
 \item The two workers receive data at constant rates. 
 \item We have a perfect estimator to accurately predict the incoming data workload on the workers. 
\end{itemize}

In Section~\ref{sec:adaptive-handling} we will relax these two assumptions. In Figure~\ref{fig:tweets-workflow}(c), the original load ratio of $J_6$ to $J_4$ is $26{\,:\,}7$. {\sf SBK} cannot handle the skew between $J_6$ and $J_4$. The approach transfers the June month to $J_4$ (Figure~\ref{fig:phases-of-mitigation}(c1)), which does not mitigate the skew. However, {\sf SBR} can redirect $\frac{9}{26}$ of the input of $J_6$ to $J_4$, which mitigates the skew by increasing the percentage load on $J_4$ to $16$ and decreasing the percentage load on $J_6$ to $17$. An example where {\sf SBK} can mitigate the skew is the case of skew between the skewed worker $J_5$ and its helper $J_2$. {\sf SBK} can transfer the processing of May to $J_2$, which brings the two workers to a similar workload. Specifically, the percentage load on $J_2$ increases to $18$ and that on $J_5$ decreases to $15$.

It should be noted that two phases do not mean that the state transfer has to be done twice necessarily. There are implementations where the state transfer during the first phase is enough and the second phase does not require another state transfer. For example, in {\sf SBR}, the state of all keys are sent to $J_4$ in the first phase, and there is no state migration needed for the second phase. 

\begin{figure}[htbp]
    \vspace{-0.1in}
	\includegraphics[width=3.3in]{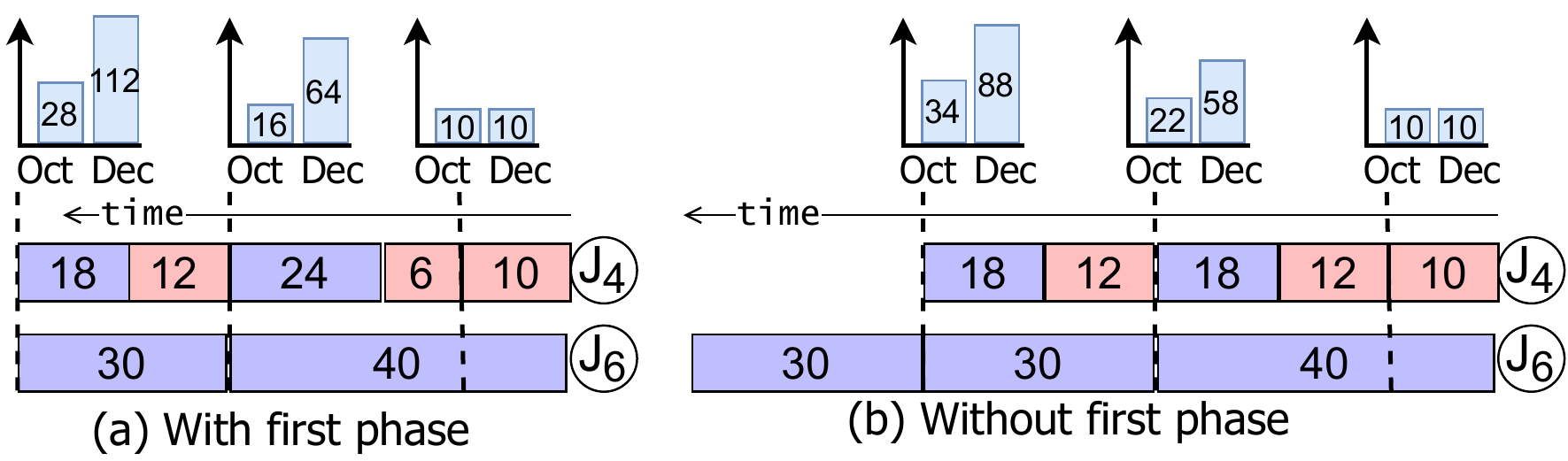}
	\vspace{-0.12in}
	\caption{\label{fig:first-phase-significance}
		\textbf{First phase helps to reflect the actual ratio of December and October tuples early. The bar charts show the progression of results as the workers process tuples}
	}
\end{figure}

\boldstart{Significance of the first phase.} \frmname has an extra phase for two reasons. First, it gives some immediate respite to the skewed worker and avoids imminent risks of the skewed worker going out of computing resources, invoking back-pressure~\cite{misc/backpressure} etc. Second, it may allow the user to see the representative results earlier compared to the case where there is only one phase. Figure~\ref{fig:first-phase-significance} illustrates this idea. For simplicity of calculation, we assume that $J_4$ processes October and $J_6$ processes December only. Notice that December tuples are almost four times the tuples of October (Figure~\ref{fig:tweets-workflow}(c)). Suppose the {\sf HashJoin} operator receives $2$ October and $8$ December tuples every second and the skew is detected when the unprocessed queue sizes of $J_4$ and $J_6$ are $10$ and $40$, respectively. Figure~\ref{fig:first-phase-significance}(a) shows the case where there exists a first phase. Suppose the first phase redirects all December tuples to $J_4$. In $3$ seconds, $J_4$ receives $24$ December and $6$ October tuples and catches up with the queue of $J_6$. After this, the second phase starts and redirects $3$ out of every $8$ December tuples to $J_4$. Assuming the workers process tuples at similar rates, the bar charts show the October and December tuples count shown to the user as the workers process more data. When the workers have processed $10$ tuples each, the bar chart shows $10$ tuples for both months. After that the effect of first phase starts. When both workers have processed $40$ tuples each, the bar chart shows $16$ tuples for October and $64$ tuples for December, which is representative of the ratio of October to December tuples in the input data. Figure~\ref{fig:first-phase-significance}(b) shows the case where there is no first phase. After detection of skew, the second phase starts and redirects $3$ out of every $8$ December tuples to $J_4$. In this case, even after both the workers have processed $40$ tuples each, the bar chart shows $22$ tuples for October and $58$ tuples for December. The ratio gradually moves towards the actual ratio of $1{\,:\,}4$ between October to December tuples.

%% file: sec4.tex
\section{Adaptive Skew Handling}
\label{sec:adaptive-handling}

In the previous section, we assumed that data arrives at constant rates to the workers and the second phase has a perfect estimator. In this section, we study the case when these assumptions are not true. In particular, variable patterns in incoming data rates and an imperfect estimator can result in erroneous workload predictions. Consequently, the second phase may not be able to keep the workload of the skewed and helper workers at a similar level. Thus, the controller may start another iteration of mitigation. Since, each iteration may incur an overhead, such as state transfer, we should try to make better workload predictions so that the number of iterations is reduced. We show that the workflow prediction accuracy depends on the skew detection threshold $\tau$ (Sections~\ref{ssec:benefit} and~\ref{ssec:tau-impact}). In order to reduce the technical burden on the user to fix an appropriate $\tau$, we develop a method to adaptively adjust $\tau$ to make better workload predictions (Section~\ref{ssec:adjust-tau-over-iterations}).




\subsection{Load reduction from mitigation}
\label{ssec:benefit}


We measure the {\em load reduction} ($LR$) from mitigation as the difference in the maximum input size received by a skewed worker and its helper without and with mitigation. Formally, let $S$ and $H$ represent the skewed worker and the helper worker, respectively. The load reduction is defined as:
\vspace{-0.07in}
\begin{equation}\label{eq:benefit-as-diff}
  LR = \left[max(\sigma^{}_S, \sigma^{}_H) \right]_{unmitigated} - \left[max(\sigma^{}_S, \sigma^{}_H) \right]_{mitigated},
\end{equation}
where $\sigma_w$ is the size of the total input received by a worker $w$ during the entire execution. 

\begin{figure}[htbp]
    \vspace{-0.1in}
	\includegraphics[width=2.8in]{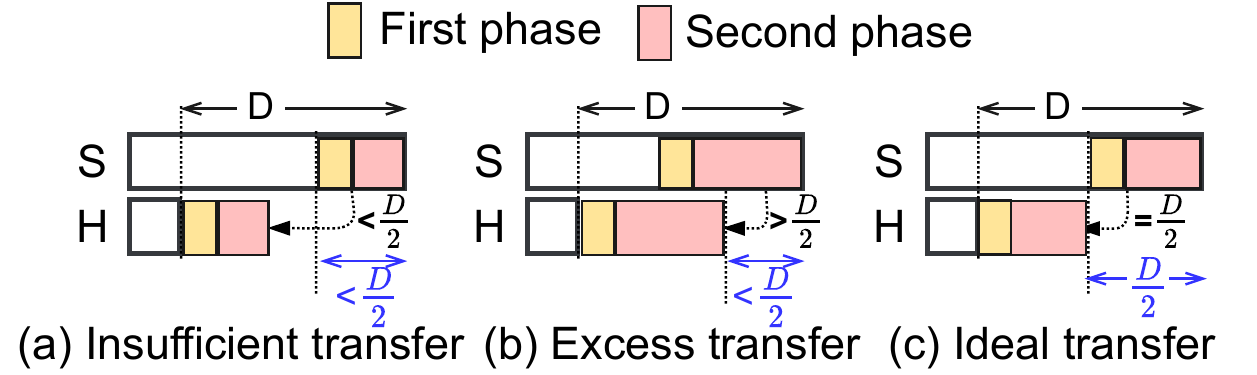} 
	\vspace{-0.12in}
	\caption{\label{fig:ideal-mitigation}
		\textbf{Effect of the amount of transferred data on the load reduction (shown in blue). The colored boxes represent the input of $S$ redirected to $H$ in the two phases.}
	}
    \vspace{-0.1in}
\end{figure}

In Figure~\ref{fig:ideal-mitigation}, $D$ represents the difference in the total input sizes of $S$ and $H$ in the unmitigated case. When mitigation is done, due to workload estimation errors, the second phase may not be able to redirect the precise amount of data to keep the workloads of $S$ and $H$ at a similar level. In Figure~\ref{fig:ideal-mitigation}(a), less than $\frac{D}{2}$ tuples of $S$ are redirected to $H$. Thus, $S$ receives more total input than $H$ and the load reduction is less than $\frac{D}{2}$. Similarly, in Figure~\ref{fig:ideal-mitigation}(b), more than $\frac{D}{2}$ tuples of $S$ are redirected to $H$. As a result, the load reduction is again less than $\frac{D}{2}$. The ideal mitigation, shown in Figure~\ref{fig:ideal-mitigation}(c), makes the total input of the two workers equal so that they finish around the same time. In particular, $\frac{D}{2}$ tuples of $S$ are sent to $H$, which is the maximum load reduction ($LR_{max}$) that can be achieved.


\vspace{-0.08in}
\subsection{Impact of $\tau$ on load reduction}
\label{ssec:tau-impact}


In this subsection, we discuss how the load reduction is affected by the value of $\tau$ at which the mitigation starts. Assume that the operator can have only one iteration of mitigation consisting of two phases. If the second phase uses a perfect estimator and the incoming data rates are constant, as assumed in Section~\ref{sec:load-transfer-mechanism}, then the maximum load reduction of $\frac{D}{2}$ can be achieved. That is:
\begin{equation}
\label{eq:ideal-total-benefit}
  LR_1 + LR_2 = \frac{D}{2},
\end{equation}
where $LR_1$ and $LR_2$ are the load reduction resulting from the first phase and second phase, respectively.



In general, the workloads estimations have errors~\cite{conf/sigmod/ChaudhuriMN98, conf/cloud/RamakrishnanSU12, journals/tpds/ChenYX15, conf/bigdataconf/YanXM13, conf/icde/GuflerARK12}. These errors can cause the second phase to redirect less or more than the ideal amount of $S$ tuples (Figure~\ref{fig:ideal-mitigation}(a,b)). In other words, the load reduction from the second phase depends on the accuracy of workload estimation.
The workload estimation accuracy depends on $\tau$ as shown next. If $\tau$ increases, then it takes a longer time for the workload difference of $S$ and $H$ to reach $\tau$, resulting in a higher sample size. Suppose the estimation accuracy increases as the sample size increases. Then a higher $\tau$ means that the system makes a more accurate workload estimation. Thus, the total load reduction can be computed as the following:
\begin{equation}\label{eq:actual-total-benefit}
  LR = LR_1 + (1-f(\tau))LR_2,
\end{equation}
where $f(\tau)$ is a function representing the error in the estimation of the future workloads. As $\tau$ increases, $f(\tau)$ decreases. 

The above analysis shows that a higher $\tau$ results in a higher load reduction. However, setting $\tau$ to an arbitrarily high value means that the system waits a long time before starting the mitigation.  Consequently, there may not be enough future input left to mitigate the skew completely. Thus the value of $\tau$ should be chosen properly to achieve a balance between a high estimation accuracy and waiting so long that the opportunity to mitigate skew is lost. This is a classic exploration-exploitation dilemma~\cite{journals/ml/AuerCF02}.

\begin{figure}[htbp]
    \vspace{-0.12in}
	\includegraphics[width=3.1in]{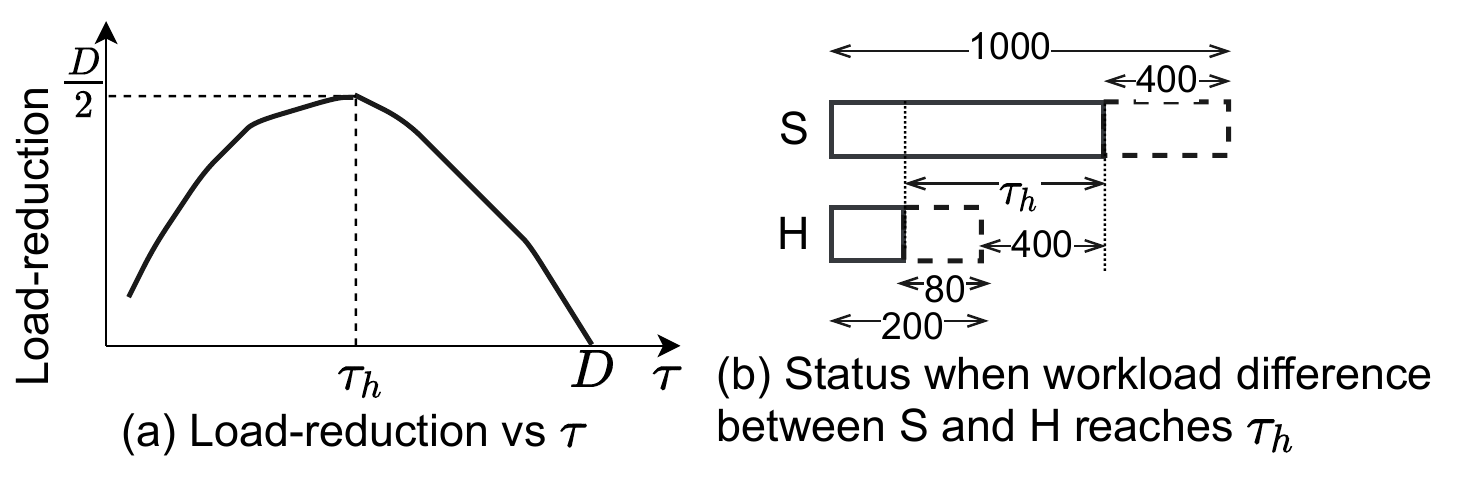} 
	\vspace{-0.12in}
	\caption{\label{fig:analysis-return}
		\textbf{Dependence of load reduction on the $\tau$.}
	}
    \vspace{-0.12in}
\end{figure}

Figure~\ref{fig:analysis-return}(a) shows the relationship between $\tau$ and load reduction. A small $\tau$ results in a small load reduction because of a high estimation error. As $\tau$ increases, $f(\tau)$ decreases and load reduction increases. The load reduction cannot exceed $LR_{max}=\frac{D}{2}$. However, the load reduction does not remain at $\frac{D}{2}$ as $\tau$ further increases, as shown next. Suppose $S$ and $H$ are to receive $1,000$ and $200$ tuples in total, respectively. Figure~\ref{fig:analysis-return}(b) shows the time when they have received $600$ and $120$ tuples respectively. At this time, the remaining $400$ tuples of $S$ can be redirected to $H$ to achieve the maximum load reduction of $\frac{D}{2}$ ($=\frac{1000-200}{2}$). The difference in the workloads of the workers at this time is denoted by $\tau_h$. After $\tau_h$, the load reduction continues to decrease because there are not enough future tuples left. Ultimately, at $\tau = D$, the load reduction becomes $0$. 


\subsection{Adaptive mitigation iterations}
\label{ssec:adjust-tau-over-iterations}

When the workloads of $S$ and $H$ diverge due to workload estimation errors, the controller may start another mitigation iteration. Section~\ref{sssec:multiple-iterations} discusses how multiple iterations of mitigation are performed. In the previous subsection, we saw that $\tau$ should be chosen appropriately to maintain a balance between workload estimation accuracy and a long delay in the start of mitigation. Section~\ref{sssec:choose-tau} shows how to autotune $\tau$ adaptively to make better workload estimations, rather than asking the user to supply an appropriate value of $\tau$.

\subsubsection{Multiple iterations of mitigation}
\label{sssec:multiple-iterations}

\begin{figure}[htbp]
    \vspace{-0.12in}
	\includegraphics[width=2.8in]{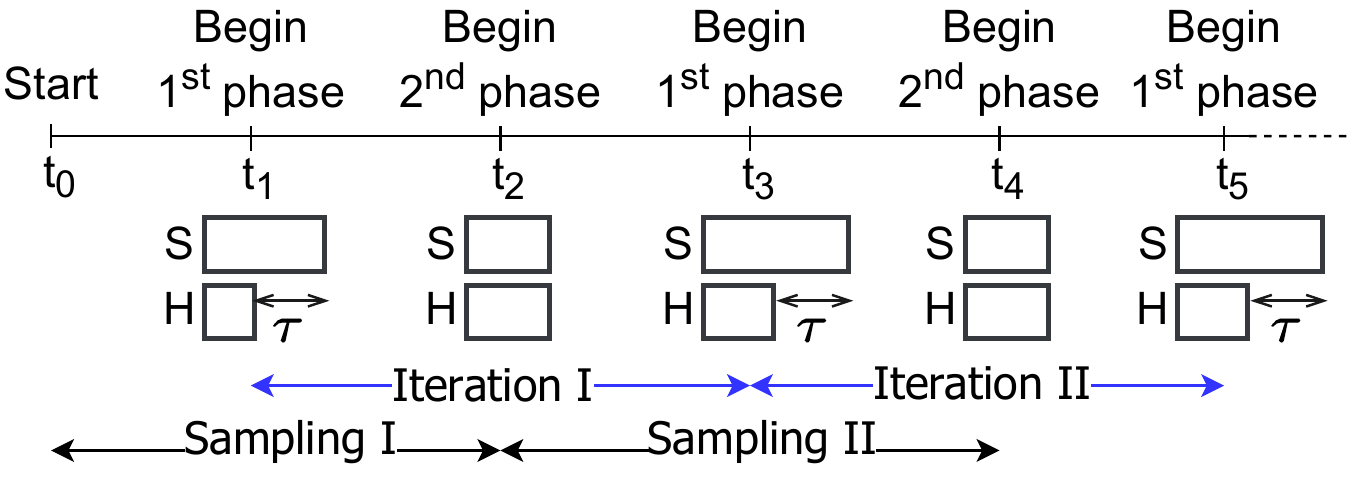}
	\vspace{-0.12in}
	\caption{\label{fig:multiple-iterations}
		\textbf{Multiple mitigation iterations}
	}
    \vspace{-0.15in}
\end{figure}

Figure~\ref{fig:multiple-iterations} shows an example timeline of two successive iterations of mitigation. The first iteration starts at $t_1$ when the difference of the workloads of $S$ and $H$ exceeds $\tau$. Their workloads are brought to a similar level at $t_2$. Then, the second phase starts. Due to workload estimation errors, the second phase redirects less than the ideal amount of tuples. Thus, the workload of $S$ gradually becomes greater than $H$. At $t_3$, their workload difference exceeds $\tau$ and the second iteration starts.

A question is how to decide the time interval from which the sample is used to do prediction~\cite{journals/tpds/XiaoSC13, conf/sc/DiKC12}. Figure~\ref{fig:multiple-iterations} shows an example that uses the sample collected since the last time when $S$ and $H$ had a similar load. Specifically, at $t_2$, the second phase of the first iteration uses the sample collected since $t_0$. The second phase of the second iteration uses the sample collected since $t_2$.

\subsubsection{Dynamically adjusting $\tau$}
\label{sssec:choose-tau}

A low value of $\tau$ causes high errors in workload estimation due to a small sample size, which in turn results in more mitigation iterations. On the other hand, a high $\tau$ may start the mitigation too late when there are not enough future tuples to mitigate the skew. Rather than using a fixed user-provided value of $\tau$, which may be too low or too high, we adaptively adjust $\tau$'s value during execution to make better workload predictions, reduce the number of iterations, and achieve higher load reduction.



In Section~\ref{sec:load-transfer-mechanism}, we introduced an estimation function $\psi$ that uses a workload sample to estimate future workloads. Let $\varepsilon$ denote the standard error of estimation~\cite{misc/prediction-interval}, which is a measure of predicted error in workload estimation. For example, the standard error for mean-model~\cite{misc/statistical-forecasting, misc/prediction-interval} estimator is $\varepsilon = d \sqrt{1 + \frac{1}{n}}$, where $d$ is the sample standard deviation and $n$ is the sample size. As mentioned in Section~\ref{ssec:tau-impact}, $\varepsilon$ decreases as $\tau$ increases.  We want $\varepsilon$ to be in a user-defined range $[\varepsilon^{}_l,\varepsilon^{}_u]$, where $\varepsilon^{}_l$ and $\varepsilon^{}_u$ are the lower and upper limits, respectively. In particular, when $\varepsilon > \varepsilon^{}_u$, we assume the error is too high and will lead to a low load reduction. Similarly, when $\varepsilon < \varepsilon^{}_l$, the error is low enough to make a good estimation.

The controller keeps track of $\varepsilon$ and adaptively adjusts $\tau$ in order to move $\varepsilon$ towards the $[\varepsilon^{}_l,\varepsilon^{}_u]$ range. Algorithm~\ref{alg:adjusting-tau} describes the process of adjusting $\tau$. For a worker $w$, let $\phi^{}_w$ represent the current workload and $\hat{\phi}^{}_w$ represent the workload predicted by $\psi$. The controller periodically collects the current workload metrics from the workers (line~\ref{alg:collect-workload}) and adds them to the existing sample (line~\ref{alg:add-to-sample}). The function $\psi$ uses the workload sample to predict future workloads and outputs $\varepsilon$ in the prediction (line~\ref{alg:estimate}). Once $\varepsilon$ is obtained, $\tau$ can be adjusted.

\begin{algorithm}[ht]
\caption{\hl{Dynamic $\tau$ adjustment by the controller.} \label{alg:adjusting-tau}}
        \KwIn{$[\varepsilon^{}_l, \varepsilon^{}_u]  \gets$ Standard error acceptable range}
        \KwIn{$\texttt{W}$: collected workloads sample}
        \KwIn{$\tau$: current threshold}
        \KwOut{Adjusted threshold}
       \SetKwFunction{AdjustTau}{adjust-threshold}

    \SetKwProg{myproc}{Procedure}{}{}
    $\phi^{}_S, \phi^{}_H \gets$ Collect current workloads of $S$ and $H$ \\ \label{alg:collect-workload}
    Add $\langle \phi^{}_S, \phi^{}_H \rangle$ to \texttt{W} \\ \label{alg:add-to-sample}
    $\hat{\phi}^{}_S, \hat{\phi}^{}_H, \varepsilon \gets $ Estimate future workloads of $S$ and $H$ using $\psi$   \\ \label{alg:estimate}
    \vspace{0.05in}
    {\tt // adjust threshold} \\
    \uIf{$\phi^{}_S - \phi^{}_H >= \tau$ \textbf{and} $\varepsilon > \varepsilon^{}_u$}{
        \tcp{Higher sample size needed to lower $\varepsilon$}
        \textbf{return} increase-threshold($\tau$)
    }
    \uElseIf{$\phi^{}_S - \phi^{}_H < \tau$ \textbf{and} $\varepsilon < \varepsilon^{}_l$}{
        \tcp{$\varepsilon$ has become quite low }
        \textbf{return} decrease-threshold($\tau$)
    }
    \uElse{
        \textbf{return} $\tau$
    }
\end{algorithm}

\boldstart{Increasing $\tau$.} The need to increase $\tau$ arises when the workers $S$ and $H$ pass the {\em skew-test} (Section~\ref{ssec:skew-detection}), but $\varepsilon > \varepsilon^{}_u$. This means that a higher sample size is needed to lower $\varepsilon$. At this point, the mitigation is started and an increased $\tau$ is chosen for the next iteration to achieve a smaller $\varepsilon$. The threshold $\tau$ should be cautiously increased so as to not set it to a very high value (Section~\ref{ssec:tau-impact}). 


\boldstart{Decreasing $\tau$.} Now consider the case where $S$ and $H$ do not pass the {\em skew-test} because their workload difference is less than $\tau$, but $\varepsilon < \varepsilon^{}_l$. This means that $\varepsilon$ is low and the sample size is big enough to yield a good accuracy. If we wait for the workload difference to reach $\tau$, there may not be enough data left to mitigate the skew. Thus, $\tau$ is decreased to the current workload difference ($\phi^{}_S - \phi^{}_H$) and mitigation starts right away, thus yielding a higher load reduction. 

%% file: sec5.tex
\section{\frmname on more operators}
\label{sec:other-operators}

Till now we used the running example of skew in the probe input of {\sf HashJoin}. A data analysis workflow can contain many
operators that are susceptible to partitioning skew such as {\sf sort} and {\sf group by}. In this section, we generalize \frmname to a broader set of operators. Specifically, we formalize the concept of ``operator state mutability'' in Section~\ref{ssec:formalize-state}. In Section~\ref{ssec:immutable-state-operators}, we discuss the impact of state mutability on state migration. In Sections~\ref{ssec:approach-1-mutable} and \ref{ssec:approach-2-mutable}, we use the load-transfer approaches described in Section~\ref{sec:load-transfer-mechanism} to handle skew in mutable-state operators. We discuss a state migration challenge when using the ``split by records'' approach and explain how to handle it.

\subsection{Mutability of operator states }
\label{ssec:formalize-state}

In this subsection, we define two types of operator states, namely {\em immutable state} and {\em mutable state}. When an operator receives input partitioned by keys, the state information of keys is stored in the operator as {\em keyed states}~\cite{journals/pvldb/CarboneEFHRT17}. Each keyed state is a mapping of type 
\vspace{-0.06in}
$$scope\rightarrow val,$$

\vspace{-0.06in}
\noindent
where $scope$ is a single key or a set or range of keys, and $val$ is information associated with the $scope$. For example, in {\sf HashJoin}, each join key is a $scope$, and the list of build tuples with the key is the corresponding $val$. Similarly, in a hash-based implementation of {\sf group-by}, each individual group is a $scope$, and the aggregated value for the group is the corresponding $val$. In a range-partitioned {\sf sort} operator, a range of keys is a $scope$, and the sorted list of tuples in the range is the corresponding $val$. In the rest of this section, for simplicity, we use the term ``state'' to refer to ``keyed state.''

An input tuple uses the state associated with the $scope$ of the key of the tuple. If the $val$ of this $scope$ cannot change, we say the state is {\em immutable}; otherwise, it is called {\em mutable}.
For example, the processing of a probe tuple in {\sf HashJoin} does not modify the list of build tuples for its key. Such operators whose states are immutable are called {\em immutable-state operators}. On the other hand, an input tuple to {\sf sort} is added to the sorted list associated with its $scope$ (range of keys), thus it modifies the state. Such operators that have a mutable state are called {\em mutable-state operators}. 

Notice that the execution of an operator can have more than one phase.  For instance, a {\sf HashJoin} operator has two phases, namely the build phase and the probe phase. The concept of mutability is with respect to a specific phase of the operator.  In {\sf HashJoin}, the states in the build phase are mutable, while the states in the probe phase are immutable.  \frmname is applicable to a specific phase, and its state migration depends on the mutability of the phase. Table~\ref{table:operator-types} shows a few examples of immutable-state and mutable-state operators.

\begin{table}[htbp]
\small{
\begin{tabular}{|p{1.75cm}|p{6.2cm}|}
\hline
\textbf{Immutable-state operator} & {\sf HashJoin}~(Probe phase), {\sf HB Set Difference}~(Probe phase), {\sf HB Set Intersection}~(Probe phase) \\ \hline
\textbf{Mutable-state operator} & {\sf HashJoin}~(Build phase), {\sf HB Group-by}, {\sf RB Sort}, {\sf HB Set Difference}~(Build phase), {\sf HB Set Intersection}~(Build phase), {\sf HB Set Union} \\ \hline 
\end{tabular}
}
\textit{}
\caption{Examples of physical operators based on state mutability. $HB$ means hash-based and $RB$ means range-based.}
\label{table:operator-types}
\vspace{-0.16in}
\end{table}

\subsection{Impact of mutability on state migration}
\label{ssec:immutable-state-operators}

\begin{figure}[htbp]
    \vspace{-0.12in}
	\includegraphics[width=0.85\linewidth]{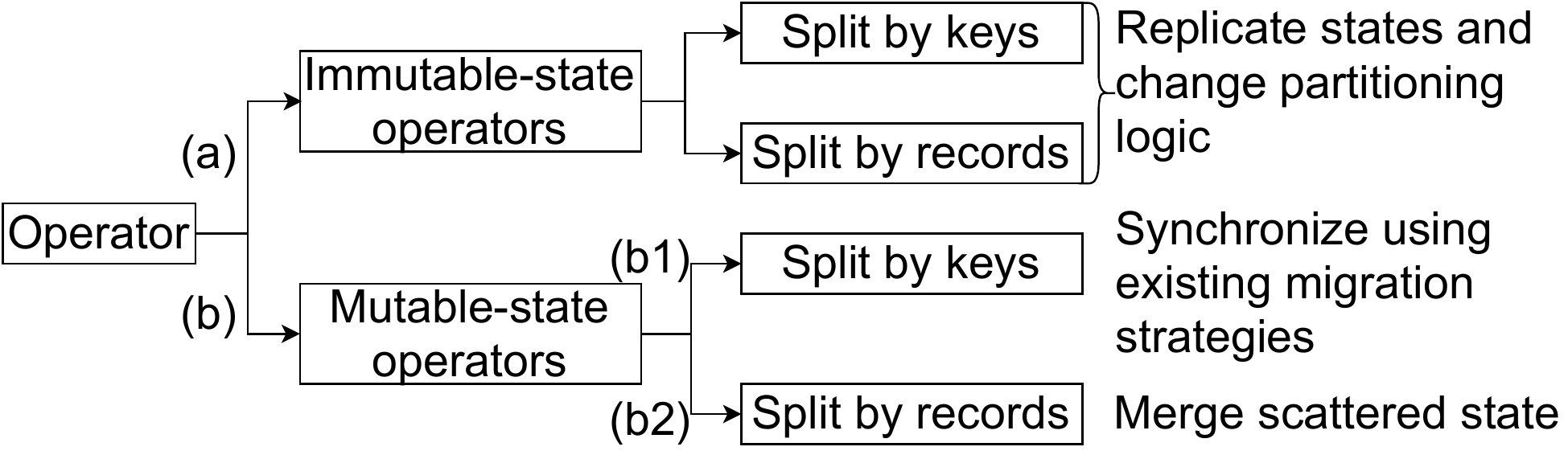} 
	\vspace{-0.12in}
	\caption{\label{fig:operator-tree}
		\textbf{Operator state mutability and state migration.}
	}
    \vspace{-0.15in}
\end{figure}

Figure~\ref{fig:operator-tree} shows how to handle state migration for operators when using the two load-transfer approaches discussed in Section~\ref{sec:load-transfer-mechanism}. 

The state-migration process for immutable-state operators, as shown in branch~(a) in Figure~\ref{fig:operator-tree}, involves replicating the skewed worker's states at the helper, followed by a change in the partitioning logic. Thus, the tuples redirected from the skewed worker to the helper can use the state of their $scope$ at the latter. In contrast, the state-migration process is more challenging for mutable-state operators (branch~(b)) because it is difficult to synchronize the state transfer and change of partitioning logic for a mutable state~\cite{journals/pvldb/MaiZPXSVCKMKDR18}. State-migration strategies that focus on such synchronization exist in the literature and will be briefly discussed in Section~\ref{ssec:approach-1-mutable}. As we show in Section~\ref{ssec:approach-2-mutable}, such a synchronization is not always possible. Next, we discuss how to do state migration when using the two load-transfer approaches in mutable-state operators.



\subsection{Mutable-state operators: split by keys}
\label{ssec:approach-1-mutable}

The {\sf SBK} approach offloads the processing of certain keys in the skewed worker partition to the helper. Consider a {\sf group-by} operator that receives covid related tweets and aggregates the count of tweets per month. The skewed worker offloads the processing of a month (say, June) to the helper. There needs to be a synchronization between state transfer and change of partitioning logic so that the redirected June tuples arriving at the helper use the state formed from all June tuples received till then. In the case of {\sf group-by}, this state is the count of all June tuples received by the operator.  Existing work on state-migration strategies focuses on this synchronization. A simple way to do this synchronization is to pause the execution, migrate the state, and then resume the execution~\cite{conf/sigmod/ArmbrustDTYZX0S18,journals/pvldb/CarboneEFHRT17,conf/icde/ShahHCF03}. A drawback of this approach is that pausing multiple times for each iteration may be a significant overhead. Another strategy is to use markers~\cite{journals/pvldb/ElseidyEVK14}. The workers of the previous operator emit markers when they change the partitioning logic. When the markers from all the previous workers are received by the skewed and helper workers, the state can be safely migrated. Thus, skew handling in mutable-state operators using the ``split by keys'' approach can be safely done by using one of these state-migration strategies (branch~(b1) in Figure~\ref{fig:operator-tree}).

\subsection{Mutable-state operators: split by records}
\label{ssec:approach-2-mutable}

\begin{figure}[htbp]
    \vspace{-0.1in}
	\includegraphics[width=0.95\linewidth]{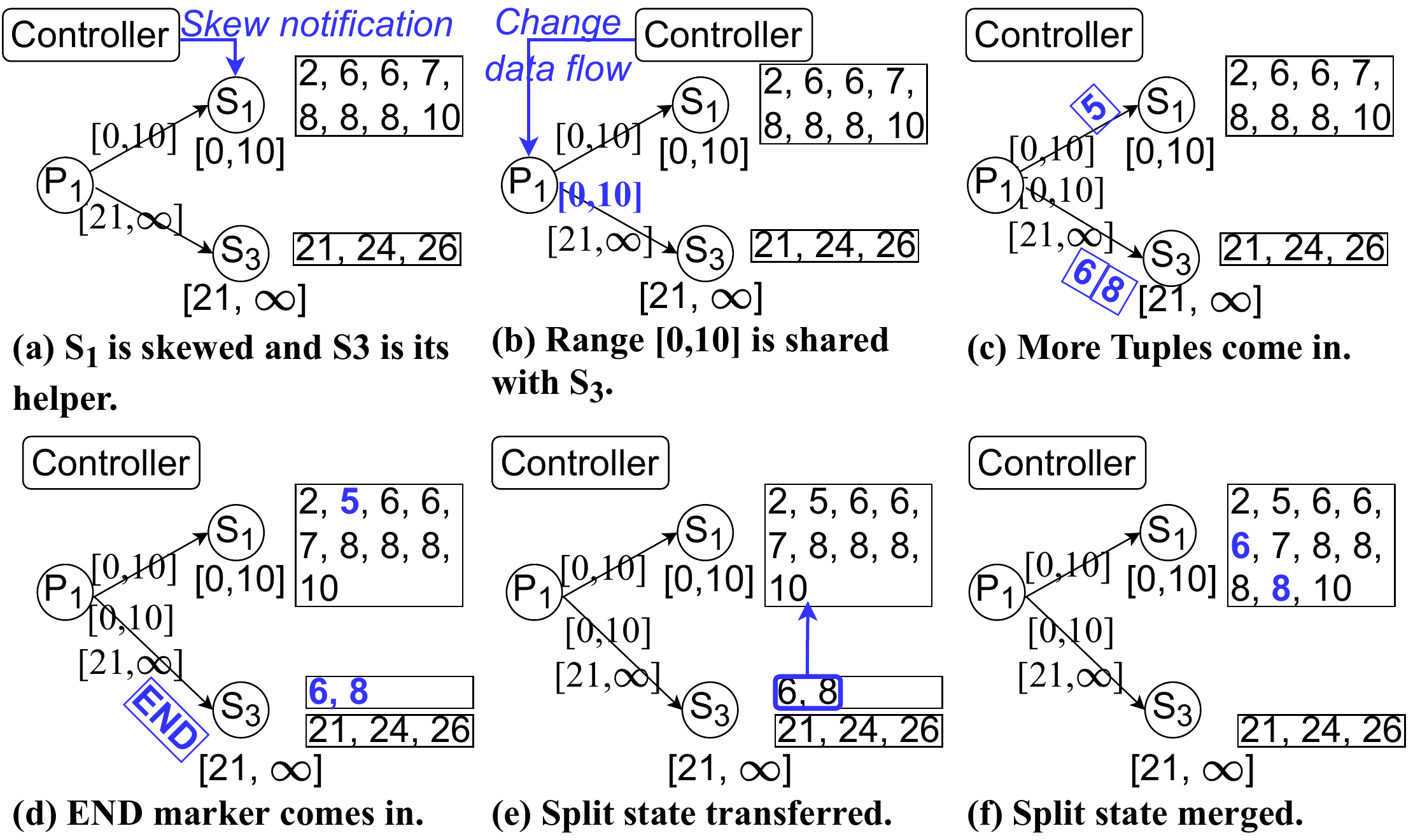} 
	\vspace{-0.12in}
	\caption{\label{fig:reshape-on-sort}
		\textbf{Skew handling using the ``split by records'' approach in the sort operator. $S_2$ is omitted for simplicity.}
	}
    \vspace{-0.15in}
\end{figure}

In this subsection, we use the {\sf SBR} approach in mutable-state operators (branch~(b2) in Figure~\ref{fig:operator-tree}). We show that the synchronization between state transfer and change of partitioning logic is not possible when using this approach and discuss its effects. Consider a sort operator with three workers, namely $S_1$, $S_2$, and $S_3$, which receive range-partitioned inputs. The ranges assigned to the three workers are $[0,10]$, $[11,20]$, and $[21,\infty]$. As shown in Figure~\ref{fig:reshape-on-sort}(a), $S_1$ is skewed and $S_3$ is its helper. The controller asks the previous operator to change its partitioning logic and send the tuples in $[0,10]$ to both $S_1$ and $S_3$ (Figures~\ref{fig:reshape-on-sort}(b,c)). The synchronization of state migration and change of partitioning logic by the aforementioned state-migration strategies relies on an assumption that, at any given time, the partitioning logic sends tuples of a particular $scope$ to a single worker only. When the tuples of $[0,10]$ are sent to both $S_1$ and $S_3$, this assumption is no longer valid. Worker $S_3$ saves the tuples from the range $[0,10]$ in a separate sorted list (Figure~\ref{fig:reshape-on-sort}(d)). Such a scenario where the $val$ of a $scope$ is split between workers is referred to as a {\em scattered state}.

This scattered state needs to be merged before outputting the results to the next operator. Now we explain a way to resolve the scattered state problem. When a worker of the previous operator finishes sending all its data, it notifies the sort workers by sending an {\sf END} marker (Figure~\ref{fig:reshape-on-sort}(d)). When $S_3$ receives {\sf END} markers from all the previous workers, it transfers its tuples in the range $[0,10]$ to the correct destination of those tuples, i.e., $S_1$ (Figure~\ref{fig:reshape-on-sort}(e,f)), thus merging the scattered states for the $[0,10]$ range. 

We specify sufficient conditions for a mutable-state operator to be able to resolve the scattered state issue. The above approach of merging the scattered parts is suited for blocking operators such as group-by and sort, which produce output only after processing all the input data. Thus, the above approach can be used by mutable-state operators if they can 1) combine the scattered parts of the state to create the final state, and 2) block outputting the results till the scattered parts of the state have been combined. 

%% file: sec6.tex
\vspace{-0.05in}
\section{\frmname in Broader Settings}
\label{sec:broader-settings}

Our discussion about \frmname so far is based on several assumptions in Section~\ref{sec:overview} for simplification. Next we relax these assumptions.

\vspace{-0.05in}
\subsection{High state-migration time}

The state-migration time is assumed to be small till now.  In this subsection, we study the case where this time could be significant. 

\boldstart{Precondition for skew mitigation.} In the discussion in Section~\ref{sec:overview}, state migration is started immediately after skew detection. If the time to migrate state is more than the time left in the execution, the state migration is futile. Thus, the controller checks if the estimated state-migration time is less than the estimated time left in the execution and only then proceeds with state migration. The state-migration time can be estimated based on factors such as state-size and serialization cost~\cite{conf/networking/YunLWRK20,journals/corr/DingFMWYZC15}. The time left in the execution can be estimated by monitoring the input data remaining to be processed and the processing speed~\cite{conf/sigmod/KwonBHR12} or by using the historical data~\cite{conf/icac/GuptaMD08}. 

\begin{figure}[htbp]
    \vspace{-0.12in}
	\includegraphics[width=0.7\linewidth]{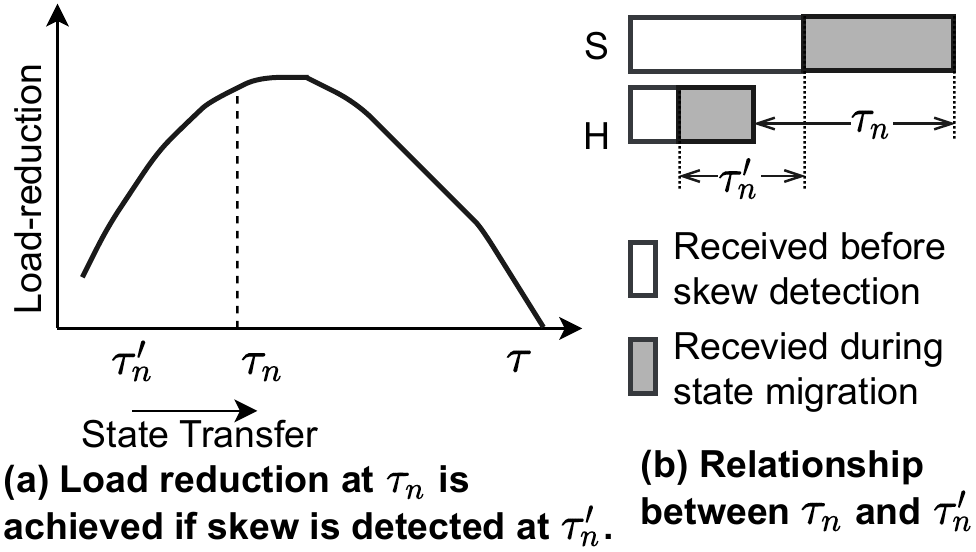} 
	\vspace{-0.12in}
	\caption{\label{fig:state-transfer-effect}
		\textbf{Adapt $\tau$ by considering the state-transfer time.}
	}
    \vspace{-0.12in}
\end{figure}

\boldstart{Dynamic adaptation of $\tau$.} Suppose the adapted value of $\tau$ output by Algorithm~\ref{alg:adjusting-tau} to be used in the next iteration is $\tau_{n}$. The discussion in Section~\ref{sssec:choose-tau} assumes that the load transfer begins when the workload difference is around $\tau_{n}$. This is possible only when the state-migration time is small. When the time is significant, the load transfer will start when the workload difference becomes considerably greater than $\tau_{n}$. In order to start the load transfer at $\tau_{n}$ (as assumed by Section~\ref{sssec:choose-tau}), the skew has to be detected earlier. Thus, we adjust the skew detection threshold to $\tau_n'$, which is less than $\tau_n$, such that the state migration starts when the workload difference is $\tau_n'$ and ends when the workload difference is $\tau_n$ (Figure~\ref{fig:state-transfer-effect}). 

Formally, suppose $t$ is the number of tuples processed by the operator per unit time, $M$ is the estimated state-migration time, and $\hat{f}_S$ and $\hat{f}_{H}$ are the predicted workload percentages of $S$ and $H$, respectively. The estimated difference in the tuples received by $S$ and $H$ during the state migration is $(\hat{f}_S - \hat{f}_H)*t*M$. Therefore, given $\tau_{n}$, the value of $\tau_{n}'$ can be calculated as follows:
\vspace{-0.03in}
$$ \tau_{n}' = \tau_{n} - (\hat{f}_S - \hat{f}_H)*t*M.$$



\subsection{Multiple helper workers}
\label{ssec:multiple-helper-workers}

Till now we have assumed a single helper per skewer worker. Next, we extend \frmname to the case of multiple helpers.



\boldstart{Load reduction.} The load reduction definition (Section~\ref{ssec:benefit}) can be extended for $S$ and its helpers $h_1,\ldots,h_n$ as follows:
$$
  LR = \max_{w \in \{S,h_1,h_2,\ldots,h_n\}}(\sigma_w) - \max_{w \in \{S,h_1,h_2,\ldots,h_n\}}(\sigma'_w).
$$
In the equation, $\sigma_w$ and $\sigma'_w$ are the sizes of the total input received by worker $w$ during the entire execution in the unmitigated case and mitigated case, respectively. Suppose $T$ is the total number of tuples received by the operator and $f_w$ is the actual workload percentage of a worker $w$. In the unmitigated case, $S$ receives the maximum total input among $S$ and its helpers, which is $f_S*T$ tuples. In the ideal mitigation case, $S$ and its helpers have the same workload, which is the average of the workloads that they would have received in the unmitigated case. As discussed in Section~\ref{ssec:benefit}, the ideal mitigation results in maximum load reduction denoted as:
$$
  LR_{max} = \big(f_S - \dfrac{\sum_{w\in \{S,h_1,h_2,\ldots,h_n\}}{f_w}}{n+1}\big) * T.
$$

\begin{figure}[htbp]
    \vspace{-0.1in}
	\includegraphics[width=0.75\linewidth]{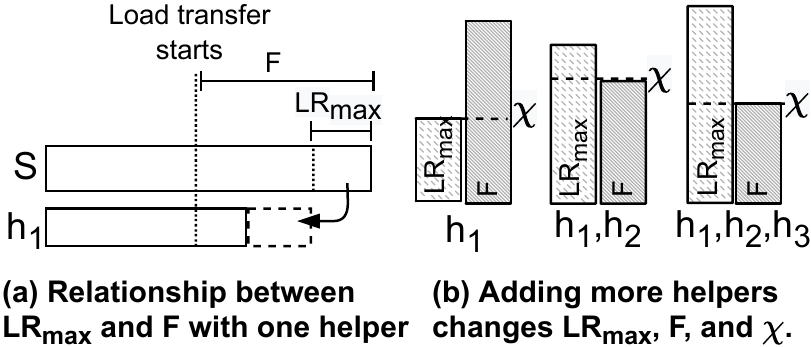} 
	\vspace{-0.12in}
	\caption{\label{fig:multiple-helpers}
		\textbf{Choosing appropriate helpers.}
	}
    \vspace{-0.15in}
\end{figure}

\boldstart{Choosing appropriate helpers.} We examine the trade-off between the load reduction and the state-migration overhead to determine an appropriate set of helpers for $S$. Let $h_1,\ldots,h_c$ be $c$ helper candidates for $S$ in the increasing order of their workloads. From the definition above, increasing the number of helpers results in a higher $LR_{max}$, provided the average workload percentage reduces. However, increasing the number of helpers may result in higher state-migration time since more data needs to be transferred. Suppose $L$ is the number of future tuples to be processed by the operator at the time of skew detection. The estimated number of future tuples left to be processed by $S$ after state migration is $F = (L-M*t)*\hat{f}_S$.
Increasing the number of helpers may increase the state-migration time ($M$) and thus decrease $F$, which means that there are fewer future tuples of $S$ to do load transfer. Thus, given a set of helpers, the highest possible load reduction after state migration is $\chi = min(LR_{max}, F)$. As we add more helpers, $\chi$ initially increases and then starts decreasing. The set of helpers chosen right before $\chi$ starts decreasing are appropriate. Figure~\ref{fig:multiple-helpers} illustrates an example. Let $W$ be the set of helper workers, which is initially empty. After adding $h_1$ to $W$, we have $LR_{max} < F$, thus $\chi = LR_{max}$. Then, we add $h_2$ to $W$, which decreases $F$, and $\chi = F$. Then, we add $h_3$ to $W$, which decreases $F$ further and causes $\chi$ to start decreasing. Hence, the final set of helpers for $S$ is $\{h_1,h_2\}$.


\subsection{Unbounded data}
\label{ssec:infinte-data}

The input has been assumed to be bounded till now. Next, we discuss a few considerations when the input is unbounded.

\boldstart{Load reduction and impact of $\tau$.} In Section~\ref{ssec:benefit}, the load reduction was calculated based on the total input received by the workers. For the unbounded case, the load reduction can be calculated based on the input received by the workers in a fixed period of time. The impact of $\tau$ on the load reduction holds for unbounded case too. A small value of $\tau$ results in high errors in workload estimation, which leads to a small load reduction. A large value of $\tau$ that takes too long to reach is not preferred in the unbounded case either. If a large $\tau$ delays mitigation, it can lead to back pressure, loss of throughput, and even crashing of data-processing pipelines. The latency of processing can increase, causing adverse effects on time-sensitive applications such as image classification in surveillance~\cite{conf/osdi/HsiehABVBPGM18}.

\boldstart{Merging scattered states.} For bounded data, the scattered states in mutable-state operators were merged after the operator processed all the input. For unbounded data, the scattered states can be merged when the operator has to output results, e.g., when a watermark is received~\cite{journals/pvldb/BegoliACHKKMS21}.

%% file: sec7.tex
\section{Experiments}
\label{sec:experiments}

In this section, we present an experimental evaluation of \frmname using real and synthetic data sets on clusters.

\subsection{Setting}
\begin{figure*}[htbp]
    \vspace{-0.12in}
	\includegraphics[width=0.95\linewidth]{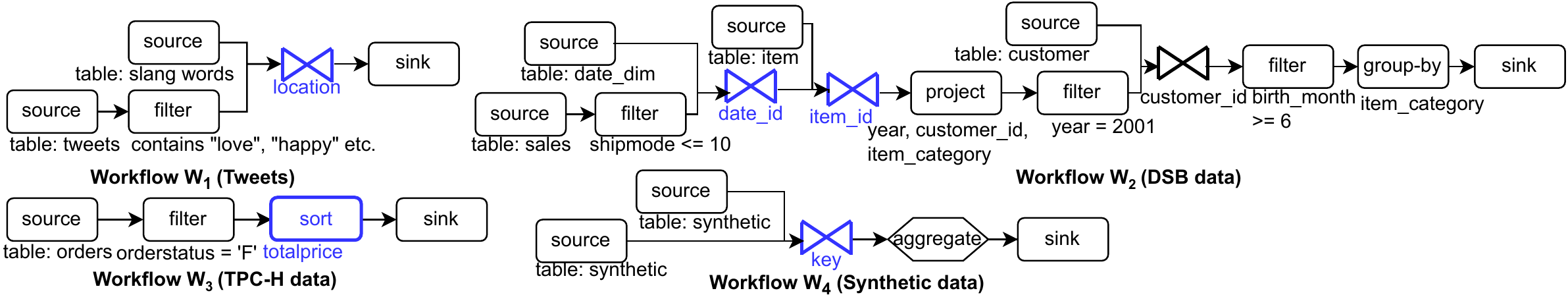} 
	\vspace{-0.12in}
	\caption{\label{fig:workflows-experiments}
		\textbf{Workflows used in the experiments. The operators with skew are shown in blue.}
	}
\end{figure*}
\noindent \textbf{Datasets and workflows.} 
We used four datasets in the experiments. The first one included $180$M tweets in the US between 2015 and 2021 collected from Twitter. The second dataset was generated using the {\sf DSB} benchmark~\cite{journals/pvldb/DingCGN21}, which is an enhanced version of {\sf TPC-DS} containing more skewed attributes, to produce record sets of different sizes ranging from $100$GB to $200$GB by varying the scaling factor. The third dataset was generated using the {\sf TPC-H} benchmark~\cite{misc/TPC-H} to produce record sets ranging from $50$GB to $200$GB. The fourth dataset was generated to simulate a changing key distribution during the execution. It included a synthetic table of $80$M tuples and another table of $4,200$ tuples, and each table had two numerical attributes representing keys and values. 

\begin{figure}[htbp]
\vspace{-0.12in}
     \centering
     \begin{subfigure}[t]{0.32\columnwidth}
          \includegraphics[width=\linewidth]{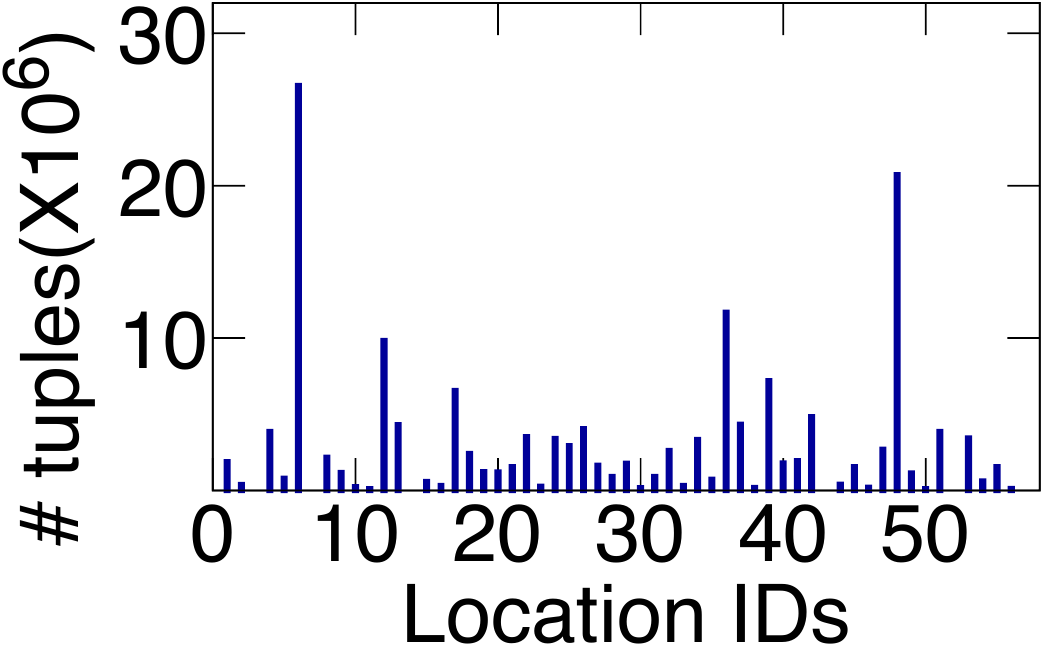}
          \caption{Tweet data.}
          \label{fig:tweet}
     \end{subfigure}
     \hfill
     \begin{subfigure}[t]{0.32\columnwidth}
          \includegraphics[width=\linewidth]{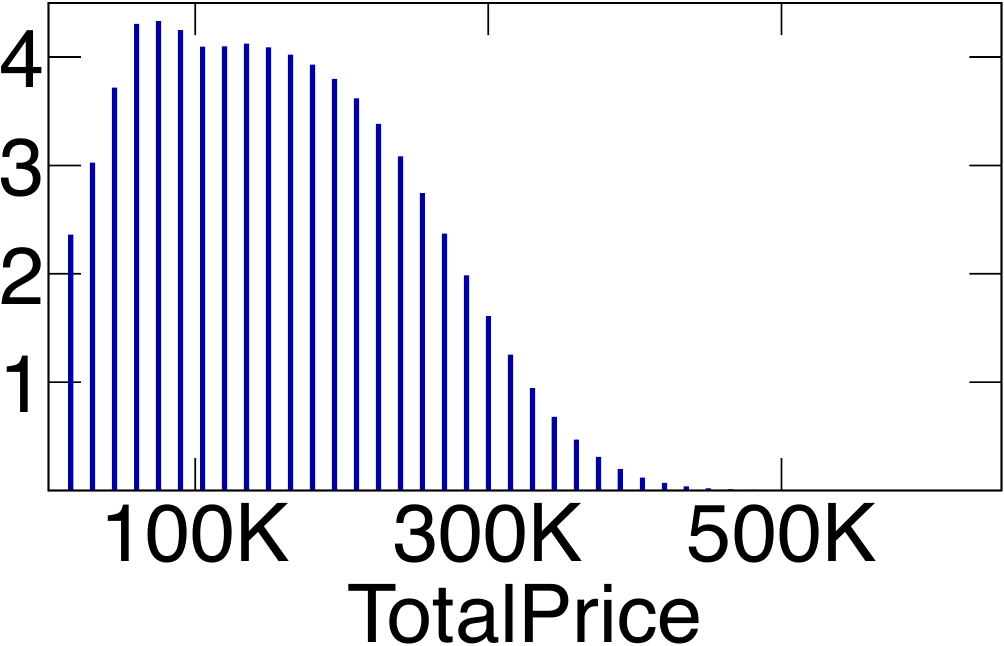}
          \caption{TPC-H data.}
          \label{fig:tpch}
     \end{subfigure}
     \begin{subfigure}[t]{0.32\columnwidth}
          \includegraphics[width=\linewidth]{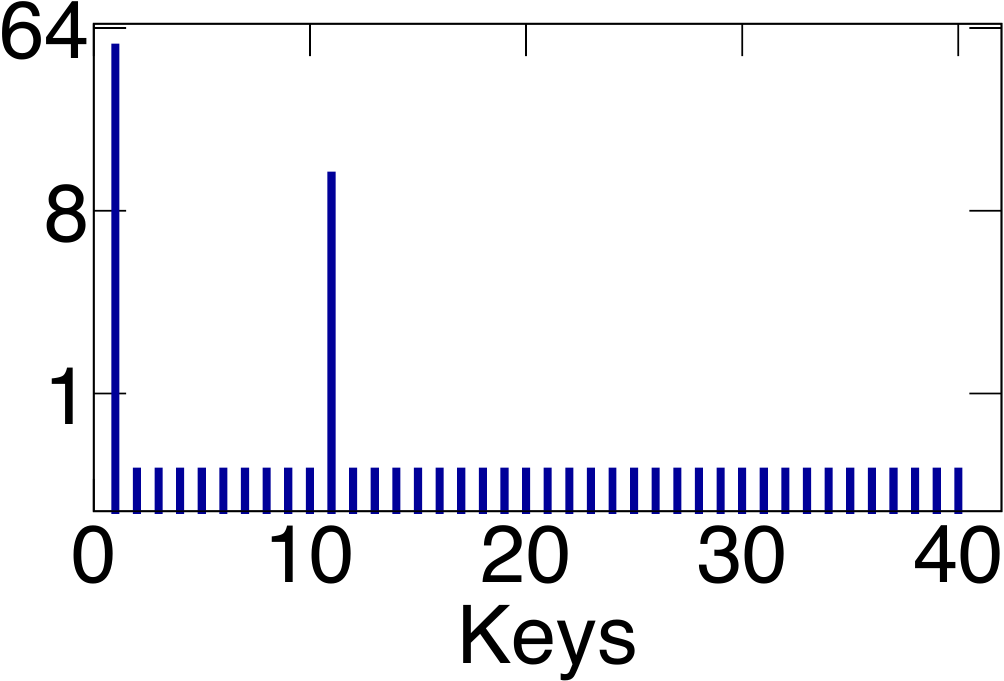}
          \caption{Synthetic data.}
          \label{fig:synthetic}
     \end{subfigure}
     \begin{subfigure}[t]{0.32\columnwidth}
          \includegraphics[width=\linewidth]{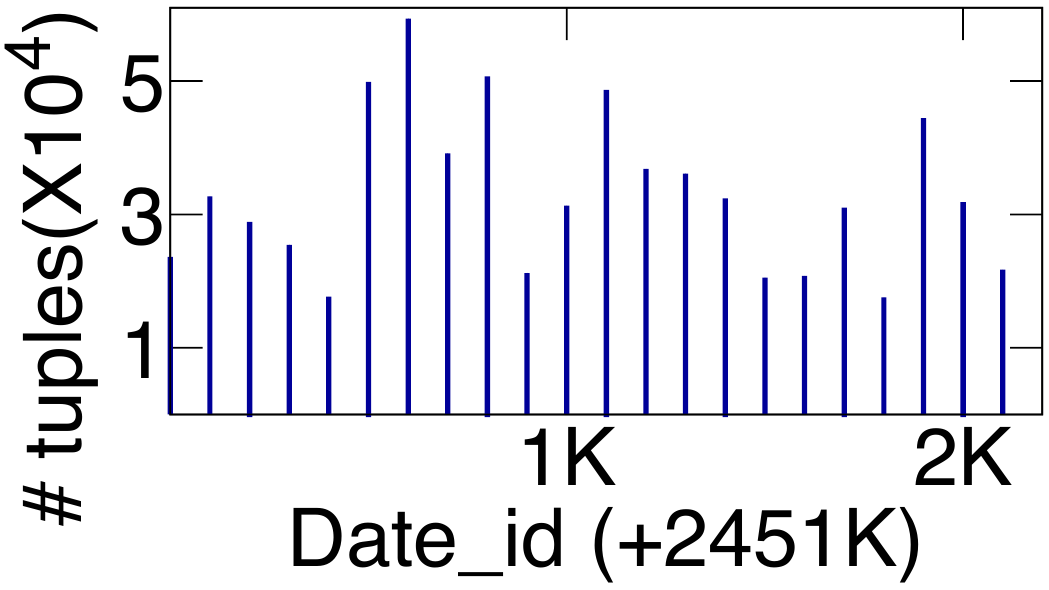}
          \caption{DSB sales data (date column).}
          \label{fig:dsb-date}
     \end{subfigure}
     \begin{subfigure}[t]{0.32\columnwidth}
          \includegraphics[width=\linewidth]{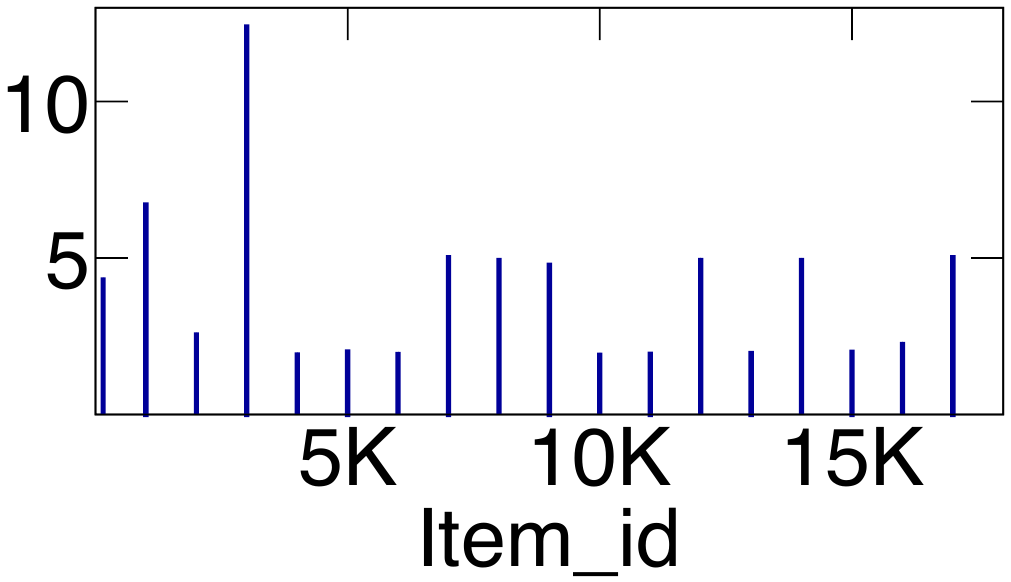}
          \caption{DSB sales data (item column).}
          \label{fig:dsb-item}
     \end{subfigure}
     \begin{subfigure}[t]{0.32\columnwidth}
          \includegraphics[width=\linewidth]{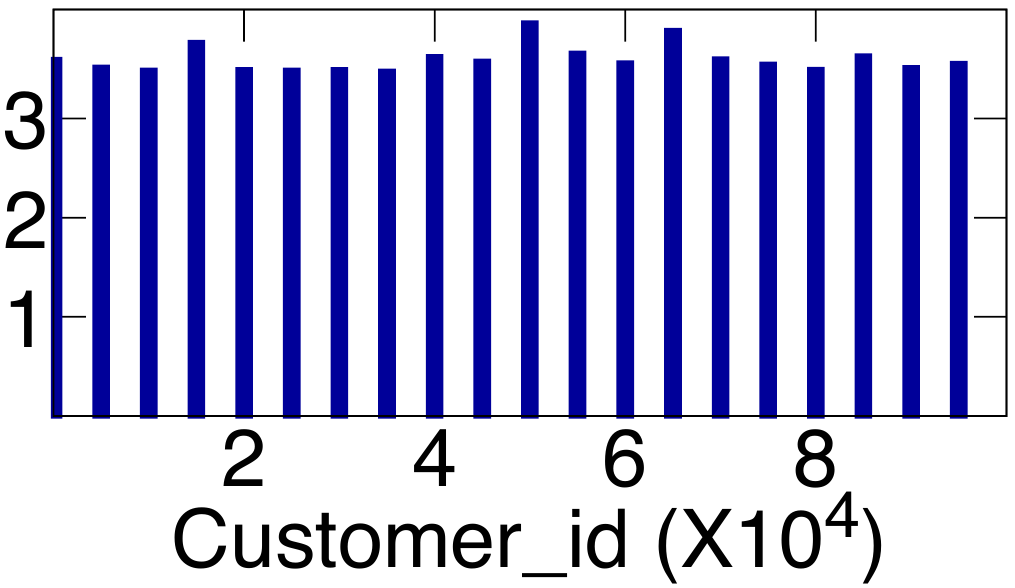}
          \caption{DSB sales data (customer column).}
          \label{fig:dsb-customer}
     \end{subfigure}
     \vspace{-0.12in}
    \caption{Partitioning-key distributions for the datasets.}
    \label{fig:key-distribution}
\vspace{-0.12in}
\end{figure}

We constructed workflows of varying complexities as shown in Figure~\ref{fig:workflows-experiments}. 
Workflow $W_1$ analyzed tweets by joining them with a table of the top slang words from the location of the tweet. This workflow is used for social media analysis to find how often people use local slang in their tweets.
The tweets were filtered on certain keywords to get tweets of a particular category. Workflow $W_2$ was constructed based on {\sf TPC-DS} query $18$, and it calculated the total count per item category for the web sales in the year $2001$ by customers whose $birth\_month>=6$. Workflow $W_3$ read the {\tt Orders} table from the {\sf TPC-H} dataset and filtered it on the {\tt orderstatus} attribute before sorting the tuples on the {\tt totalprice} attribute.  Workflow $W_4$ joined the two synthetic tables on the key attribute. Figure~\ref{fig:key-distribution} shows the distribution of the datasets that may cause skew in the workflows. Figure~\ref{fig:tweet} shows the frequency of tweets, used in $W_1$, based on the location attribute. Figure~\ref{fig:tpch} shows the distribution of the {\tt Orders} table on its {\tt totalprice} attribute, used in $W_3$, for a $100$GB TPC-H dataset. Figure~\ref{fig:synthetic} shows the distribution of the larger synthetic table in $W_4$ on the key attribute. Figures~\ref{fig:dsb-date}-\ref{fig:dsb-customer} show the distribution of the three attributes of the sales table in $W_2$ used in the three join operations for a $1$GB dataset.

\boldstart{Reshape implementation.} We implemented \frmname\footnote{\frmname is available on Github (\url{https://github.com/Reshape-skew-handle}).} on top of two open source engines, namely {\sf Amber}~\cite{journals/pvldb/KumarWNL20} and Apache {\sf Flink} (release 1.13).  In Amber, we used its native API to implement the control messages used in \frmname. Unless otherwise stated, we set both $\tau$ and $\eta$ to $100$. We used the mean model~\cite{misc/statistical-forecasting} to predict the workload of workers. In Flink, we used the {\em busyTimeMsPerSecond} metric of each task, which is the time ratio for a task to be busy, to determine the load on a task. We leveraged the mailbox of tasks (workers) to enable the control messages to change partitioning logic.  The control messages are sent to the mailbox of a task, and these messages are processed with a higher priority than data messages in a different channel. Using these control messages, we implemented the two phases of the {\sf SBR} load transfer approach on Flink as discussed in Section~\ref{sec:load-transfer-mechanism}.


\boldstart{Baselines.} For comparison purposes, we also implemented {\sf Flow-Join} and {\sf Flux} on Amber with a few adaptations. For {\sf Flow-Join}, we used a fixed time duration at the start to find the overloaded keys. The workload on a worker was measured by its input queue size. For {\sf Flow-Join}, after skew is detected, the tuples of the overloaded keys are shared with the helper worker in a round-robin manner. For {\sf Flux}, after skew is detected, the processing of an appropriate set of keys is transferred from the skewed worker to its helper. For both \frmname and the baselines,  one helper worker was assigned per skewed worker, unless otherwise stated. Also, unless otherwise stated, {\sf Flux} used a $2$ second initial duration to detect overloaded keys. To be fair, when running \frmname, we also had an initial delay of $2$ seconds to start gathering metrics and subsequent skew handling by \frmname.

All experiments were conducted on the Google Cloud Platform (GCP). The data was stored in an HDFS file system on a GCP dataproc storage cluster of 6 e2-highmem-4 machines, each with 4 vCPU's, 32 GB memory, and a 500GB HDD. 
The workflow execution was on a separate processing cluster of e2-highmem-4 machines with a 100GB HDD, running the Ubuntu 18.04.5 LTS operating system. In all the experiments, one machine was used to only run the controller. We only report the number of data-processing machines. The number of workers per operator was equal to the total number of cores in the data-processing machines and the workers were equally distributed among the machines.

\subsection{Effect on results shown to the user}
\label{ssec:results-produced}
\begin{figure}[htbp]
    \vspace{-0.1in}
	\includegraphics[width=0.6\linewidth]{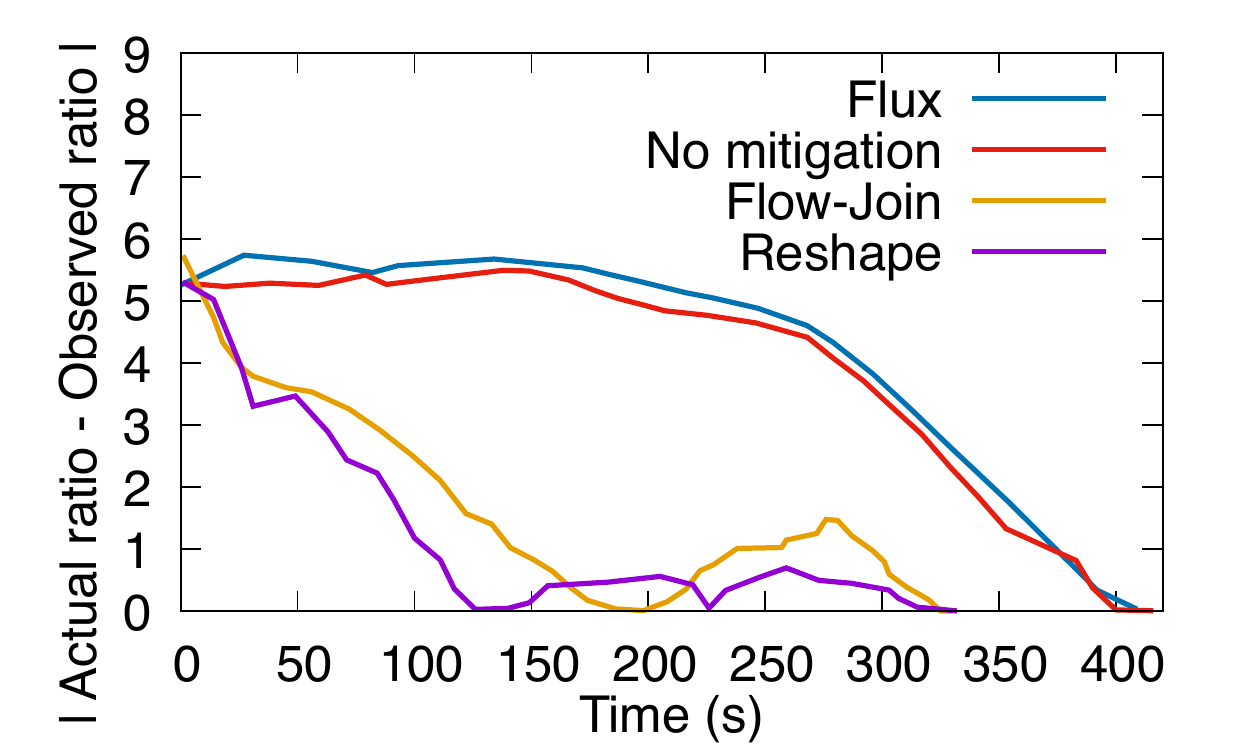} 
	\vspace{-0.12in}
	\caption{\label{fig:tweet-ratio-az}
	\textbf{Effect of the mitigation strategies on the ratio of CA to AZ tweets.}}
	\vspace{-0.15in}
\end{figure}

\begin{figure}[htbp]
    \vspace{-0.1in}
	\includegraphics[width=0.6\linewidth]{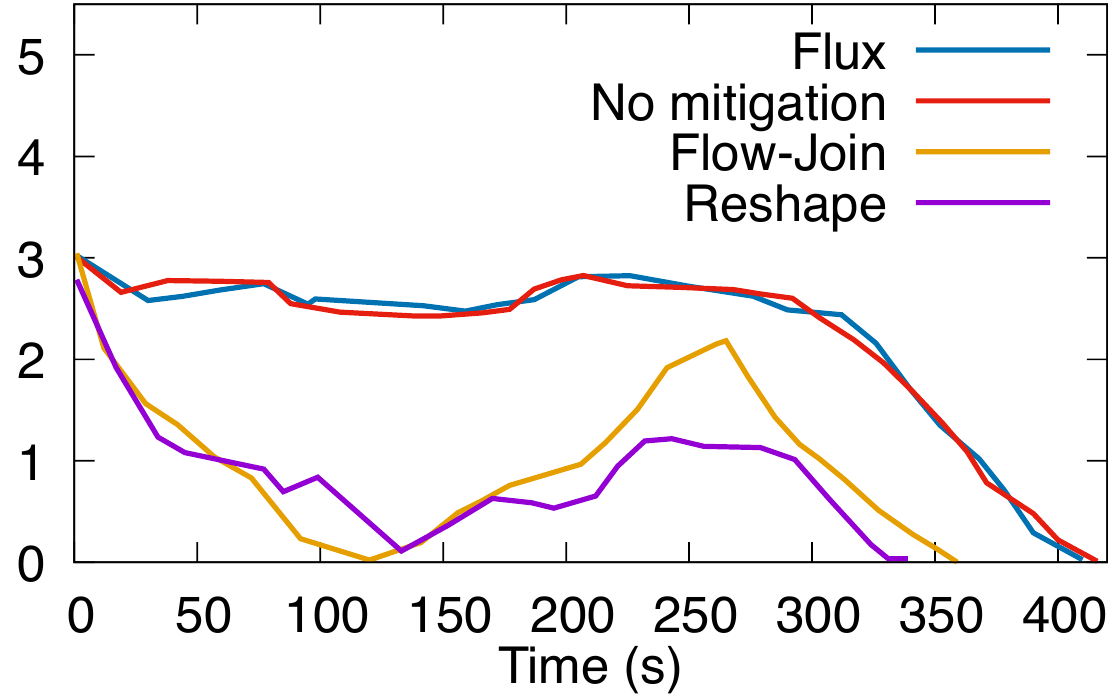} 
	\vspace{-0.12in}
	\caption{\label{fig:tweet-ratio-il}
	\textbf{Effect of the mitigation strategies on the ratio of CA to IL tweets.}}
	\vspace{-0.15in}
\end{figure}

We evaluated the effect of skew and the different mitigation strategies on the results shown to the user. We ran the experiment on $48$ cores ($12$ machines). California (location $6$) produced the highest number of tweets ($26$M) in the tweet dataset. Arizona (location $4$) and Illinois (location $17$) produced $3.8$M and $6.5$M tweets, respectively. In the unmitigated case, the tuples of California (CA), Arizona (AZ), and Illinois (IL) were processed by workers $6$, $4$, and $17$, respectively. We performed two sets of experiments, in which we mitigated the load on worker $6$ processing CA tweets by using different helper workers. In the first set of experiments, we used worker $4$ as the helper and monitored the ratio of CA to AZ tweets processed by the join operator. In the second set, we used worker $17$ as the helper and monitored the ratio of CA to IL tweets processed by the join operator. The line charts in Figure~\ref{fig:tweet-ratio-az} and \ref{fig:tweet-ratio-il} show the absolute difference of the observed ratio from the actual ratio as execution progressed. In the tweet dataset, the actual ratio of CA to AZ tweets was $6.85$ and CA to IL tweets was $4.05$. 


{\bf No mitigation}: When there was no mitigation, the CA, AZ, and IL tweets were processed at a similar rate as explained in Section~\ref{ssec:two-load-transfer-approaches}. The observed ratio remained close to $1$ till worker $4$ was about to finish processing AZ tweets in Figure~\ref{fig:tweet-ratio-az} and worker $17$ was about to finish IL tweets in Figure~\ref{fig:tweet-ratio-il}. The observed ratio started to increase (absolute difference of observed ratio with actual ratio started to decrease) after that because worker $6$ continued to process CA tweets. The actual ratio was observed near the end of execution (about $416$ seconds) in the unmitigated case.

{\bf Flux}: It used the {\sf SBK} load-transfer approach. It had the limitation of not being able to split the processing of a single key over multiple workers. The skewed worker $6$, apart from CA, was also processing the tweets from West Virginia. The processing of the tweets from West Virginia (about $600$K) was moved to the helper worker by Flux. However, this did affect the observed ratio of tweets much.

{\bf Flow-Join}: It used the {\sf SBR} approach. The execution finished earlier because the approach mitigated the skew in worker $6$. {\sf Flow-Join} had two drawbacks. First, it did not perform mitigation iteratively. It changed its partitioning logic only once based on the heavy hitters detected initially. Second, it did not consider the loads on the helper and the skewed worker while deciding the portion of the skewed worker's load to be transferred to the helper. It always transferred $50$\% of the load of the skewed worker to the helper. The observed ratio of tweets started increasing once skew mitigation started. It reached the actual ratio $198$ seconds in Figure~\ref{fig:tweet-ratio-az} and around $120$ seconds in Figure~\ref{fig:tweet-ratio-il}. Due to the aforementioned drawbacks, the observed ratio of tweets continued to increase even after reaching the actual ratio because the skewed worker continued to transfer 50\% of its load to the helper. The observed ratio continued to increase till it reached about $8.3$ in Figure~\ref{fig:tweet-ratio-az} (absolute difference = $1.5$) and $6.2$ in Figure~\ref{fig:tweet-ratio-il} (absolute difference = $2.1$). At this point, the execution was near its end and the ratio started to decrease to the actual final ratio. 

{\bf Reshape}: It used the {\sf SBR} approach and could split the processing of the CA key with a helper worker. \frmname had the advantage of iteratively adapting its partitioning logic and considered the current loads on the helper and the skewed worker while deciding the portion of load to be transferred in the second phase (Section~\ref{ssec:two-phases}). Thus, \frmname kept the workload of the skewed worker and the helper at similar levels. In Figure~\ref{fig:tweet-ratio-az} and \ref{fig:tweet-ratio-il}, after  the observed ratio reaches the actual ratio at about $120$ seconds and $130$ seconds, respectively, \frmname kept the observed ratio near the actual ratio.

\subsection{Benefits of the first phase}
\label{ssec:first-phase-benefit}

\begin{figure}[htbp]
    \vspace{-0.1in}
	\includegraphics[width=0.6\linewidth]{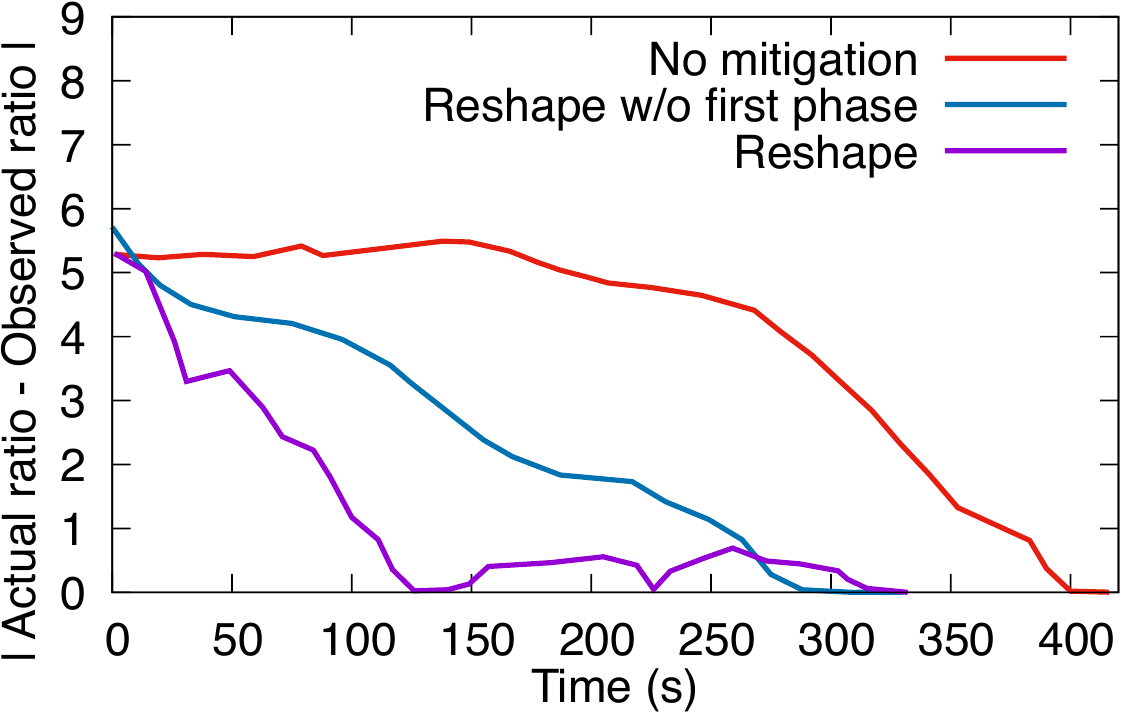}
	\vspace{-0.12in}
	\caption{\label{fig:tweet-ratio-az-first-phase-benefit}
	\textbf{Effect of first phase on the ratio of CA to AZ tweets.}}
	\vspace{-0.15in}
\end{figure}

\begin{figure}[htbp]
    \vspace{-0.1in}
	\includegraphics[width=0.6\linewidth]{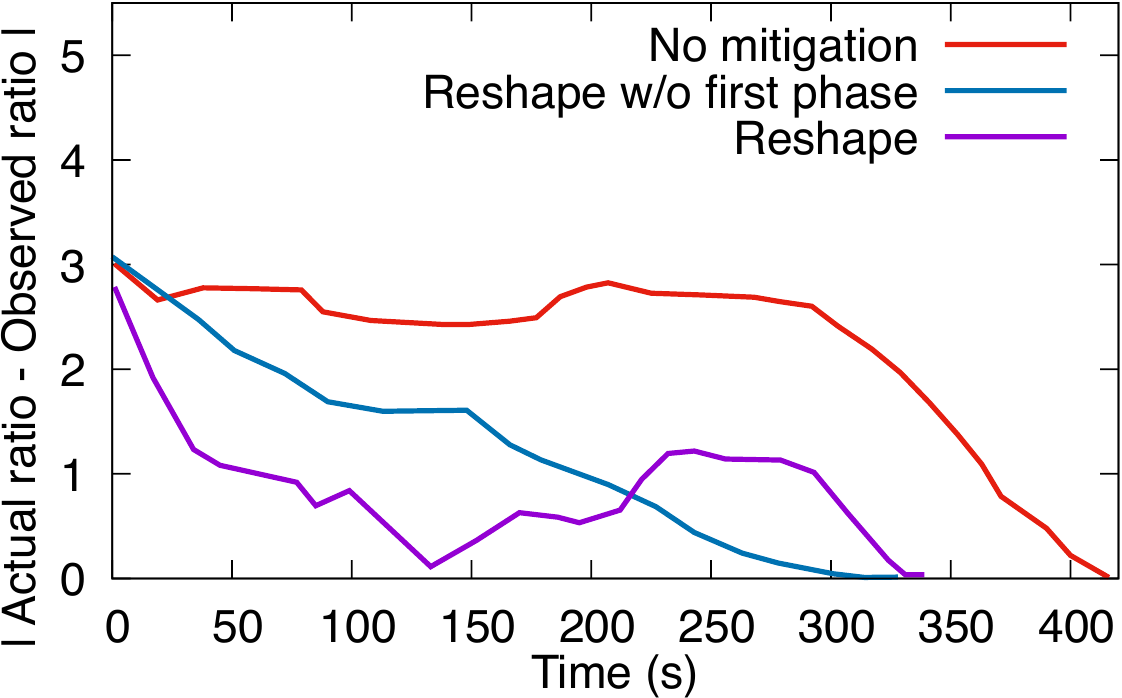}
	\vspace{-0.12in}
	\caption{\label{fig:tweet-ratio-il-first-phase-benefit}
	\textbf{Effect of first phase on the ratio of CA to IL tweets.}}
	\vspace{-0.15in}
\end{figure}

We evaluated the benefits of the first phase in Reshape as discussed in Section~\ref{ssec:two-phases}. We followed a similar setting as in the experiment in Section~\ref{ssec:results-produced} to monitor the ratio of processed tweets. There were two mitigation strategies used in this experiment. The first one was normal Reshape, with the two phases of load transfer. In the second strategy, we disabled the first phase in Reshape and just did load transfer using the second phase. The results are plotted in Figure~\ref{fig:tweet-ratio-az-first-phase-benefit} and \ref{fig:tweet-ratio-il-first-phase-benefit}.

The first phase quickly removed the existing imbalance of load between the skewed and the helper worker when skew was detected. When the first phase was present, \frmname reached the actual ratio around $120$ and $130$ seconds in Figures~\ref{fig:tweet-ratio-az-first-phase-benefit} and \ref{fig:tweet-ratio-il-first-phase-benefit}, respectively. When the first phase was disabled, \frmname reached the actual ratio around $288$ and $310$ seconds in Figures~\ref{fig:tweet-ratio-az-first-phase-benefit} and \ref{fig:tweet-ratio-il-first-phase-benefit}, respectively. Thus, the first phase allowed \frmname to show representative results earlier. Both strategies showed more representative results than the unmitigated case.

\subsection{Effect of heavy-hitter keys}
\label{ssec:heavy-hitter}

California (location $6$) produced the highest number of tweets ($26$M) and was a heavy-hitter key in the tweet dataset. We present the results for the mitigation of the skewed worker that processed the California key.

\textbf{Load balancing ratio.} The load balancing ratio at a moment during the execution is calculated by obtaining the total counts of tuples allotted to the skewed worker and its helper till that moment, and dividing the smaller value by the larger value. We periodically recorded multiple load balancing ratios during an execution and calculated their average to get the average load balancing ratio for an execution. A higher ratio is better because it represents a more balanced workload between the skewed worker and its helper.

\begin{figure}[htbp]
    \vspace{-0.1in}
	\includegraphics[width=0.6\linewidth]{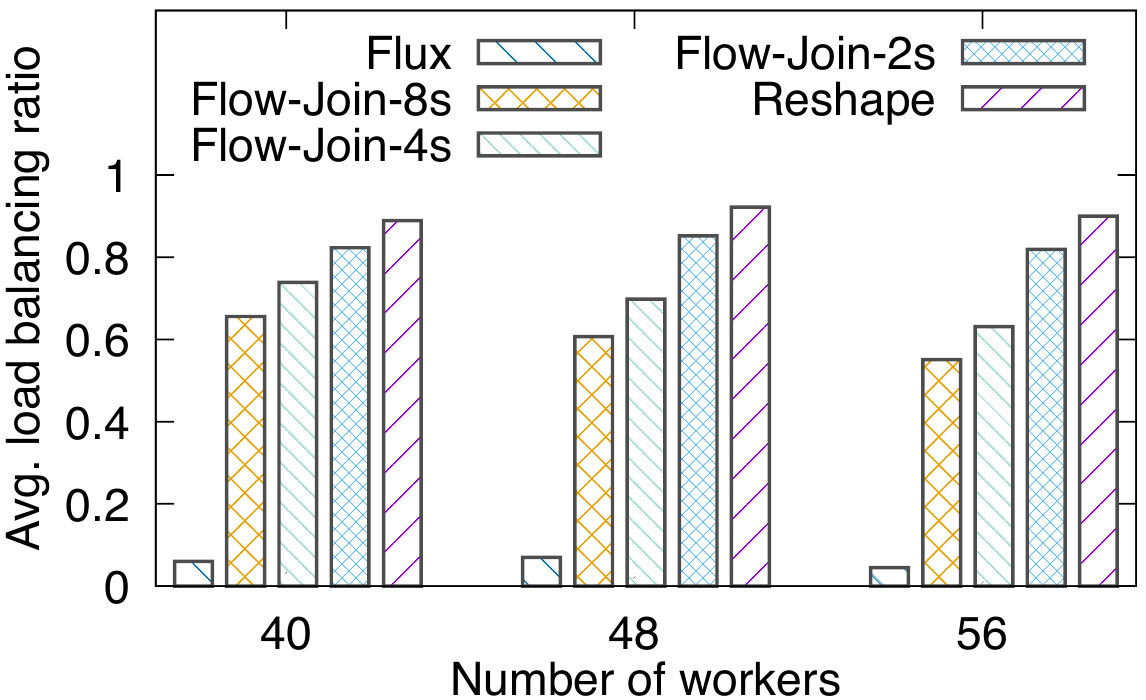} 
	\vspace{-0.12in}
	\caption{\label{fig:baseline}
	\textbf{Evaluating different methods of handling heavy-hitter keys in $W_1$ using tweets. The three {\sf Flow-Join} bars correspond to the initial delay of $2$, $4$, and $8$ seconds.}}
	\vspace{-0.15in}
\end{figure}

The average load balancing ratio for the skewed worker that processed the California key and its helpers is plotted in Figure~\ref{fig:baseline}. A higher ratio is better because it represents a more balanced workload between the skewed worker and its helper worker. We ran the experiments on three settings by varying the number of cores up to $56$ (on $14$ machines), which was the total number of distinct locations.

{\bf Flux}: It had the limitation of not being able to split the processing of a single key over multiple workers. Thus, the skewed worker processed the entire California input. The skewed worker was also processing another key with only a few hundred thousand tuples, which was moved to the helper when skew was detected. Flux had a low average load balancing ratio of about $0.06$.

{\bf Flow-Join}:  Its main drawback was the inability to do mitigation iteratively. It changed its partitioning logic once based on the heavy-hitters detected initially. The longer it spent to detect heavy-hitters with a higher confidence, the less was the amount of future tuples left to be mitigated for finite datasets. We varied the initial duration used by {\sf Flow-Join} to detect heavy-hitters from $2$ seconds to $8$ seconds. When the initial time spent was $2$ seconds, the average load balancing ratio was about $0.85$ and the final counts of tuples processed by the skewed and helper workers were approximately $14$M and $12$M, respectively. On the other hand, when the duration was $8$ seconds, the ratio was about $0.6$ and the final counts were approximately $17$M and $9$M, respectively. {\sf Flow-Join} was able to reduce the execution time of $W_1$ on $48$ cores from $416$ seconds to $302$ seconds, when the initial detection duration was $2$ seconds.

\textbf{Reshape}: It split the processing of the California key with a helper worker. \frmname had the advantage of iteratively changing its partitioning logic according to input distribution using fast control messages. Thus, the skewed and helper workers ended up processing almost similar amounts of data and the average load balancing ratio was about $0.92$. The execution time was reduced by 27\%. In particular, \frmname was able to reduce the execution time from $416$ seconds to $302$ seconds, by mitigating the skew in $W_1$ running on $48$ cores.


\subsection{Effect of latency of control messages}
\label{ssec:control-latency}

\begin{figure}[htbp]
    \vspace{-0.12in}
     \centering
     \begin{subfigure}[t]{0.35\columnwidth}
  \includegraphics[width=\linewidth]{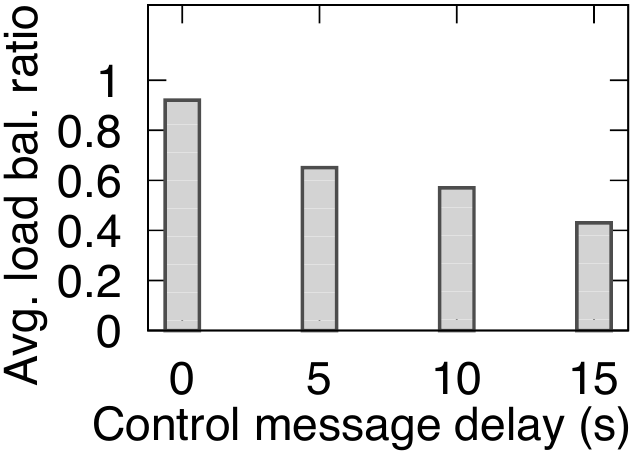}
  \caption{California data.}
  \label{fig:control-message-ca}
     \end{subfigure}
     \hspace{0.1\columnwidth}
     \begin{subfigure}[t]{0.35\columnwidth}
  \includegraphics[width=\linewidth]{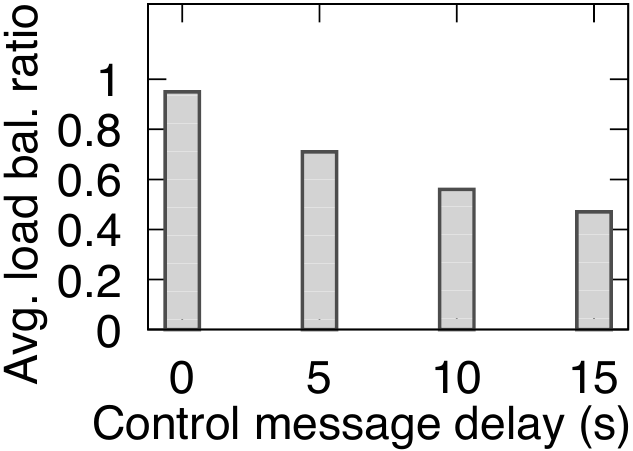}
  \caption{Texas data.}
  \label{fig:control-message-tx}
     \end{subfigure}
     \vspace{-0.13in}
        \caption{Effect of control message delay ($W_1$ on tweets).}
        \label{fig:control-message}
    \vspace{-0.12in}
\end{figure}

To evaluate the effect of the latency of control messages on skew handling by \frmname, we purposely added a delay between the time a worker receives a control message and the time it processes the message. Figure~\ref{fig:control-message} shows the result of varying the simulated delay from $0$ second (i.e., the message is processed immediately) to $15$ seconds on the mitigation of $W_1$ on $48$ cores. The figure shows the average load balancing ratio for the two pairs of skewed and helper workers processing the locations of California (location $6$) and Texas (location $48$), which had the highest counts of tweets.

{\bf Impact on responsiveness of mitigation:} As the control message delivery became slower, the delay between the controller sending a message and the resulting change in partitioning logic increased. Consider the example where the controller detected a workload difference of $350$ between the skewed worker and the helper worker and sent a message to start the first phase. In the case of no delay in control message delivery, the helper worker reached a similar workload as the skewed worker within $10$ seconds. In case of a delayed delivery, the workload difference continued to increase and got larger than $350$ before the first phase was started. For example, when there was a $5$-second delay, the workload difference was at $300$ after $10$ seconds of sending the message.


{\bf Impact on load balancing.} The latency in control messages affected the load sharing between skewed and helper workers. In the case of no delay, the two workers had almost similar loads and the average load balancing ratio was about $0.94$ as shown in Figures~\ref{fig:control-message-ca} and \ref{fig:control-message-tx}. As the delay increased, the framework was slow to react to the skew between workers, which resulted in imbalanced load-sharing. In the case of a $15$-second delay, the average load balancing ratio reduced to about $0.45$. Thus, low-latency control messages facilitated load balancing between a skewed worker and its helper.

\subsection{Benefit of dynamically adjusting $\tau$}
\label{ssec:dynamic-tau-exp}

\begin{figure}[htbp]
    \vspace{-0.1in}
	\includegraphics[width=0.6\linewidth]{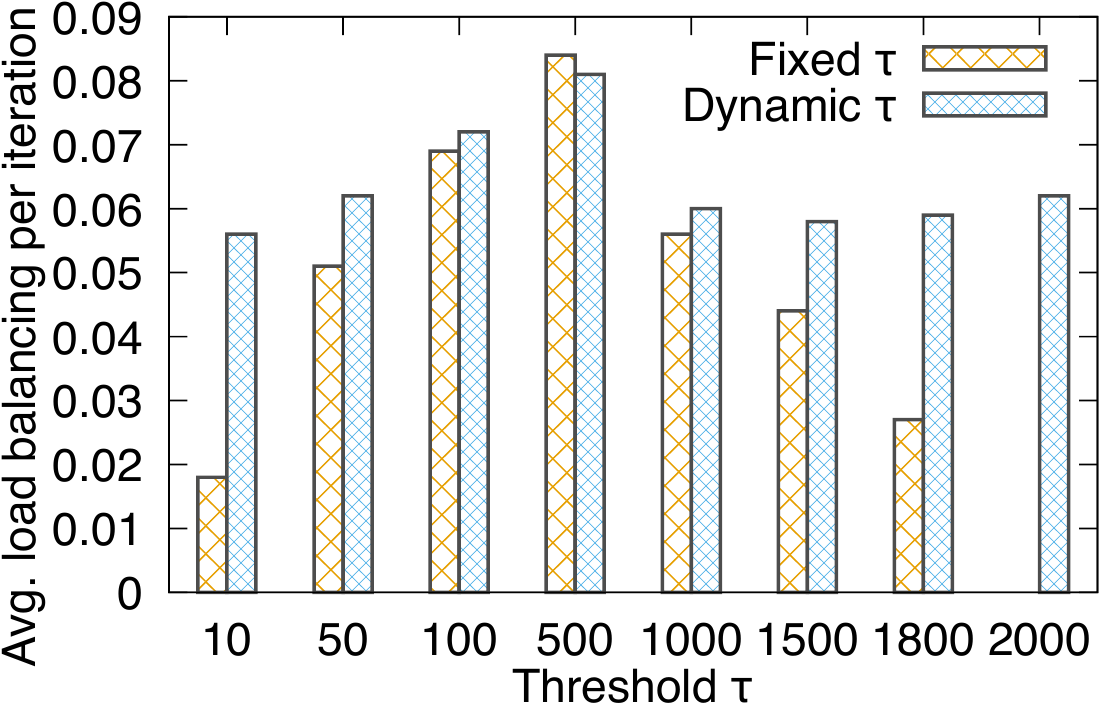} 
	\vspace{-0.12in}
	\caption{\label{fig:dynamic-tau}
	\textbf{Benefit of dynamically adjusting $\tau$ ($W_1$ on tweets).}}
	\vspace{-0.12in}
\end{figure}

We evaluated the effect of the dynamic adjustment of $\tau$ on skew mitigation in $W_1$ by \frmname on $48$ cores. We chose different values of $\tau$ ranging from $10$ to $2,000$, and did experiments for two settings. In the first setting, $\tau$ was fixed for the entire execution. In the second setting, $\tau$ was dynamically adjusted during the execution. The mean model estimated the workload of a worker as its expected number of tuples in the next $2,000$ tuples and the preferred range of standard error (Section~\ref{sssec:choose-tau}) was set to $98$ to $110$ tuples. We allowed up to three adjustments during an execution. Whenever $\tau$ had to be increased, it was increased by a fixed value of $50$. 
We calculated the average load balancing ratios for the workers processing the California and Texas keys and divided them by the total number of mitigation iterations during the execution. This resulted in the metric of average load balancing per iteration, shown in Figure~\ref{fig:dynamic-tau}. A higher value of this metric is better because it represents a more balanced workload of skewed and helper workers in fewer iterations. 

Let us first consider the cases where $\tau$ was dynamically adjusted to an increased value. Setting $\tau$ to a small value of $10$ resulted in a large number of iterations, i.e, $41$, in the fixed $\tau$ setting. In the dynamic $\tau$ setting, the controller observed that the standard error at the beginning of the second phase was greater than $110$ and increased $\tau$. Consequently, the number of iterations decreased to $14$, which resulted in a substantial increase in the metric of average load balancing per iteration. For the cases of $\tau = 50$ and $100$ in the fixed setting, the average load balancing per iteration increased with $\tau$ because the number of iterations decreased. The dynamic setting slightly decreased the iteration count in these cases.

Now let us consider the case where $\tau$ remained unchanged or decreased as a result of dynamic adjustment. When $\tau = 500$, the standard error was in the range $[98,110]$. Thus, the dynamic adjustment did not change $\tau$. When $\tau = 1000$ in the fixed setting, the mitigation started late and the workload of skewed and helper workers were not balanced. The mitigation was delayed even more for $\tau = 1500$ and $1800$ in the fixed setting and the mitigation did not happen for $\tau=2000$. In the dynamic setting for the cases of $\tau = 1000$, $1500$, $1800$, and $2000$, the controller observed that the standard error went below $98$ when the workload difference was about $700$. Thus, the controller reduced $\tau$ to $700$. The advantage of dynamically reducing $\tau$ was that it automatically started mitigation at an appropriate $\tau$, even if the initial $\tau$ was very high.

\subsection{Effect of different levels of skew}
\label{ssec:complex-workflow}

We evaluated the load balancing achieved by \frmname for different levels of skew. We used $W_2$ for this purpose. The data distributions in Figures~\ref{fig:dsb-date}-\ref{fig:dsb-item} show that the join on {\sf item\_id} was highly skewed and the join on {\sf date\_id} was moderately skewed. We evaluated the load balancing achieved for these two join operators. We scaled the data size from $100$GB to $200$GB. Meanwhile, we scaled the number of cores from $40$ to $80$ and did the experiments in each configuration.

\begin{figure}[htbp]
    \vspace{-0.12in}
	\includegraphics[width=0.6\linewidth]{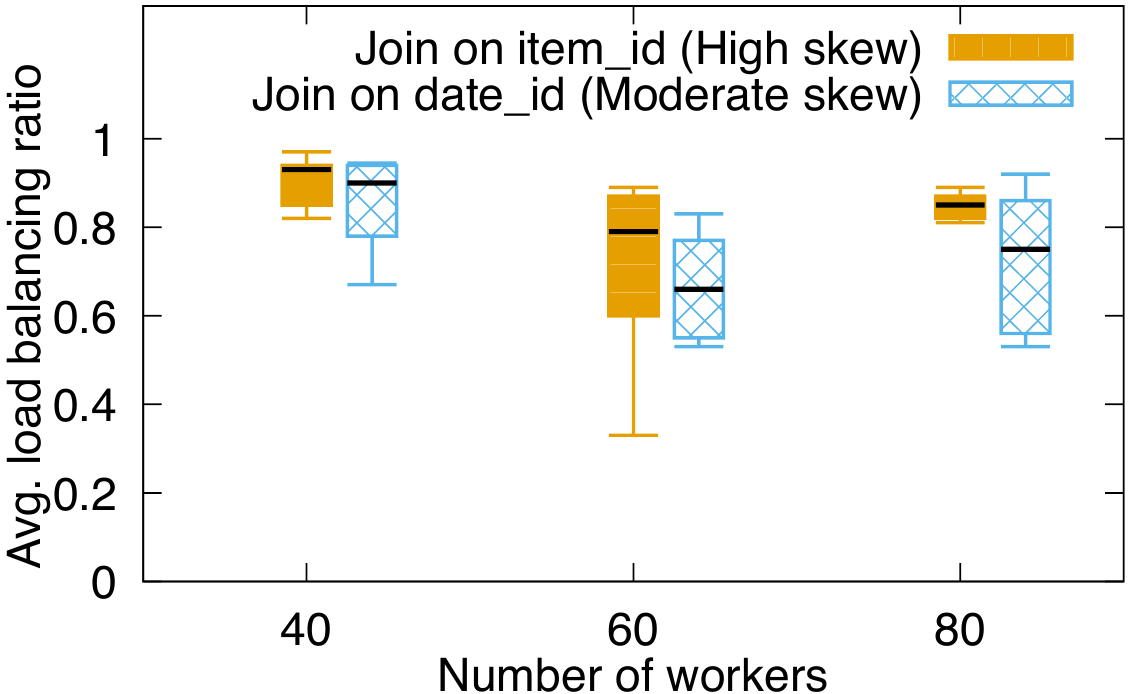} 
	\vspace{-0.12in}
	\caption{\label{fig:different-skew-levels}
	\textbf{Effect of different levels of skew ($W_2$ on DSB data). Each candlestick body represents the $25^{th}$ to $75^{th}$ percentile.}}
	\vspace{-0.15in}
\end{figure}

Figure~\ref{fig:different-skew-levels} shows the candlestick charts of the average load balancing ratios for the top five skewed workers from each of the two joins. For the highly skewed join on {\sf item\_id}, the skew was detected early, and there was enough time to transfer the load of the skewed workers to the helper workers. The $25^{th}$ and $75^{th}$ percentiles of the average load balancing ratios remained above $0.6$ for all the configurations. The median of the ratios was more than $0.77$. This result shows that \frmname was able to mitigate the skew and maintain comparable workloads on the skewed and helper workers when both the input and processing power were scaled up. The join on {\sf date\_id} had only a moderate skew, which resulted in a delayed detection of a few of its skewed workers. Due to the delayed detection, there were fewer future tuples of skewed workers to be transferred to the helpers. Thus the ratios for the join on {\sf date\_id} were lower than that for the join on {\sf item\_id}. The performance of \frmname was also shown by the reduction in the execution time. Specifically, in the case of $40$ cores, the mitigation reduced the execution time of $W_2$ from $267$ seconds to $243$ seconds. In the case of $80$ cores, the mitigation reduced the time from $335$ seconds to $269$ seconds.

\subsection{Effect of changes in input distribution}
\label{ssec:changing-data-distribution}

\begin{figure}[htbp]
	\includegraphics[width=0.6\linewidth]{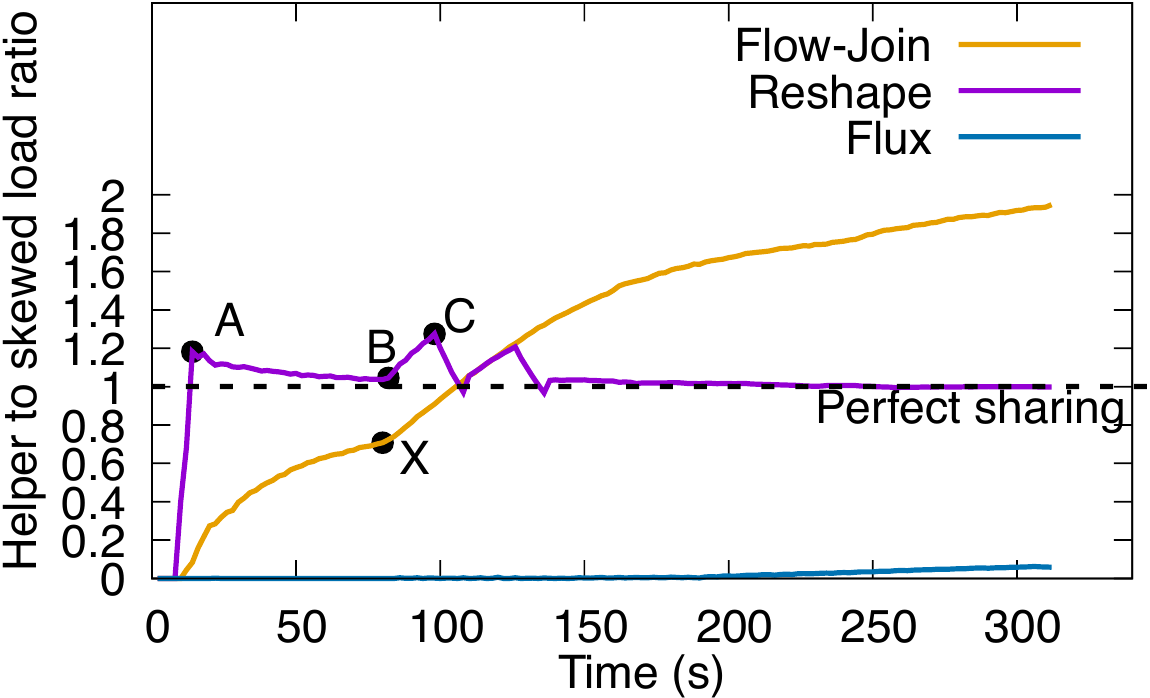} 
	\vspace{-0.12in}
	\caption{\label{fig:dynamic-distribution}
	\textbf{Effect of changes in input data distribution on load sharing ($W_4$ on the synthetic dataset).}}
	\vspace{-0.12in}
\end{figure}

We evaluated how load sharing was affected when the input distribution changed during the execution. We used the synthetic dataset and workflow $W_4$ running on $40$ cores. Both tables in the dataset had $42$ keys.  The first table contained $4,200$ tuples uniformally distributed across the keys. The second table contained $80$M tuples and was produced by the {\sf source} operator at runtime. We fixed worker $0$ and worker $10$ as the skewed and helper worker, respectively. We altered the load on key $0$ and $10$, which were processed by worker $0$ and $10$ respectively. Specifically, for the first $20$M tuples, $80$\% was allotted to the key $0$ and the rest $20$\% was uniformally distributed among the remaining keys. For the next $60$M tuples, $60$\% was allotted to the key $0$, $20$\% to key $10$, and the rest was uniformally distributed. Figure~\ref{fig:dynamic-distribution} shows the ratio of the workloads of the helper worker $10$ to the skewed worker $0$ as time progressed. We used $\tau=2,000$ to clearly show the effects of changing distributions.

\textbf{Flux.} The skewed worker was processing keys $0$ and $40$. {\sf Flux} had the limitation of not being able to split the processing of a single key over multiple workers. Upon detecting skew, {\sf Flux} can only move the key with smaller load (key $40$) to the helper. Thus, the workload ratio of helper to skewed worker remained close to $0$.

\textbf{Flow-Join:} We used a $2$-second initial duration to detect the overloaded keys. {\sf Flow-Join} identified key $0$ as overloaded and started to transfer half of its future tuples to the helper. Thus, the workload of the helper began to rise. At $80$ seconds (point {\tt X}), the input distribution changed. Since {\sf Flow-Join} cannot do mitigation iteratively, half of the tuples of key $0$ continued to be sent to the helper. The helper worker started receiving $50$\% ($=60\%*0.5 + 20\%$) and the skewed worker started receiving $30$\% ($=60\%*0.5$) of the input. Thus, the load on the helper rose and became more than the skewed worker.

\textbf{Reshape:} It started the first phase to let the helper worker quickly catch up with the skewed worker. Thus, the load of the helper sharply increased initially. After that the second phase started and the workload ratio got closer to $1$. At $80$ seconds (point {\tt B}), the input distribution changed. At point (point {\tt C}), \frmname started another iteration of mitigation and adjusted the partitioning logic according to the new input distribution. As a result, the ratio of the workloads of the workers remained close to $1$.

\subsection{Metric-collection overhead}
\label{ssec:metric-collection-overhead}

\begin{figure}[htbp]
    \vspace{-0.1in}
	\includegraphics[width=0.55\linewidth]{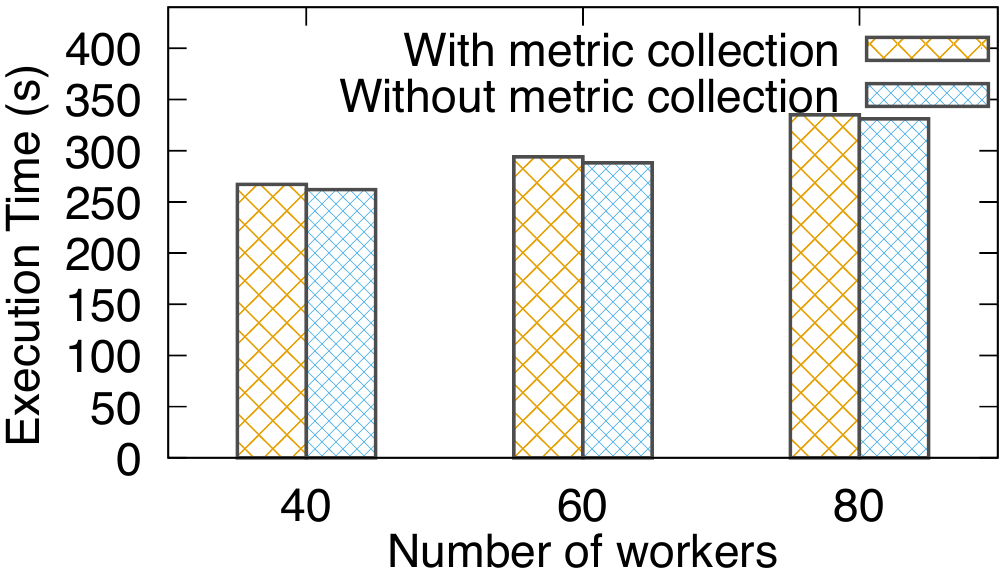} 
	\vspace{-0.12in}
	\caption{\label{fig:exp-metric-collection}
	\textbf{Metric-collection overhead ($W_2$ on DSB data).}}
	\vspace{-0.12in}
\end{figure}

We evaluated the metric-collection overhead of \frmname on the workflow $W_2$. We scaled the data size from $100$GB to $200$GB. Meanwhile, we scaled the number of cores from $40$ (on $10$ machines) to $80$ (on $20$ machines) and did the experiments in each configuration. We disabled skew mitigation and executed $W_2$ with and without metric collection to record the metric-collection overhead. As shown in Figure~\ref{fig:exp-metric-collection}, the overhead was around $1$-$2$\% for all the configurations.

\subsection{Performance of \frmname on {\sf sort}}
\label{ssec:exp-other-operators}

To evaluate its generality to other operators, we implemented \frmname for the sort operator. We used the workflow $W_3$ for this experiment. The {\em Orders} table was range-partitioned on its {\tt totalPrice} attribute. Table~\ref{table:exp-sort-reshape} lists the various percentile values of the average load balancing ratio for the skewed workers that received more than $3.5$M tuples in the unmitigated case (Figure~\ref{fig:tpch}). We scaled the data size and number of cores simultaneously from $50$GB on $20$ cores to $200$GB on $80$ cores, and did the experiment in each configuration.


\begin{table}[htbp]
\small{
\begin{tabular}{|@{ }p{1.4cm}|@{ }p{1cm}|@{ }p{1cm}|@{ }p{1cm}|@{ }p{1cm}|@{ }p{1cm}|}
\hline
{\bf \# workers} & {\bf $P_1$} & \textbf{$P_{25}$} & \textbf{$P_{50}$} & \textbf{$P_{75}$} & \textbf{$P_{99}$} \\ \hline
20 & 0.90 & 0.92 & 0.93 & 0.935 & 0.95 \\ \hline
40 & 0.84 & 0.87 & 0.89 & 0.90 & 0.91 \\ \hline
60 & 0.83 & 0.85 & 0.90 & 0.91 & 0.92 \\ \hline
80 & 0.83 & 0.84 & 0.86 & 0.87 & 0.90 \\ \hline
\end{tabular}
}
\textit{}
\caption{Average load balancing ratios when \frmname is applied on {\sf sort} ($W_3$ using the TPC-H data).}
\vspace{-0.2in}
\label{table:exp-sort-reshape}
\end{table}

As the number of cores increased, the $25^{th}$ and $75^{th}$ percentiles of the average load balancing ratios remained close to $0.9$. This result shows that the skewed and helper workers had balanced workloads when both the input and processing power were scaled up. The consistent performance of \frmname was also shown by about $20$\% reduction in the execution time. Specifically, in the case of $20$ cores, the time reduced from $789$ seconds to $643$ seconds. In the case of $80$ cores, the time reduced from $809$ seconds to $667$ seconds. 


\subsection{Effect of multiple helper workers}
\label{ssec:exp-multile-helpers}

We evaluated the load reduction achieved when multiple helper workers are assigned to a skewed worker. The experiment was done on $W_1$ running on $48$ cores. The most skewed worker among the $48$ workers received about $27$M tuples in the unmitigated case. We allotted different numbers of helpers to the skewed worker and calculated the load reduction. We set the build hash-table in each worker to have $10,000$ keys, so that the state size became significant and the state-migration time was noticeable.

\begin{figure}[htbp]
    \vspace{-0.1in}
	\includegraphics[width=0.5\linewidth]{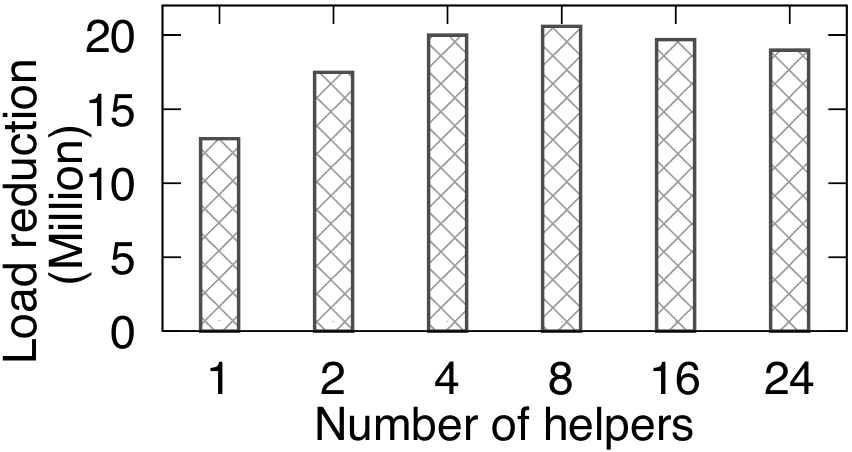} 
	\vspace{-0.12in}
	\caption{\label{fig:exp-multiple-helpers}
	\textbf{Effect of multiple helper workers ($W_1$ on tweets).}}
	\vspace{-0.12in}
\end{figure}

The results are plotted in Figure~\ref{fig:exp-multiple-helpers}. When a single helper was used, the state migration happened in $17$ seconds. The skewed worker transferred about half of its total workload to the helper, resulting in a load reduction of $13$M tuples. When $2$ helpers were used, the skewed worker transferred about two thirds of its tuples to the two helpers (about $9$M each). With more helpers, the state-migration time also increased. For $8$ helpers, the state-migration time was about $26$ seconds. Thus, there were fewer future tuples left, which resulted in a small increase in the load reduction. For $16$ helpers, the state-migration time became $32$ seconds and the load reduction decreased to $19.7$M. For $24$ helpers, the state-migration time was $39$ seconds and the load reduction decreased to $19$M.

\subsection{Performance of \frmname on Flink}
\label{ssec:flink-exp}

\begin{figure}[htbp]
     \centering
     \begin{subfigure}[t]{0.35\columnwidth}
  \includegraphics[width=\linewidth]{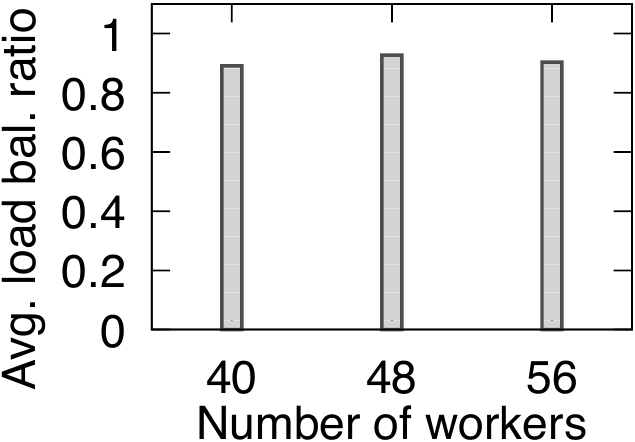}
  \caption{California data.}
  \label{fig:flink-ca}
     \end{subfigure}
     \hspace{0.1\columnwidth}
     \begin{subfigure}[t]{0.35\columnwidth}
  \includegraphics[width=\linewidth]{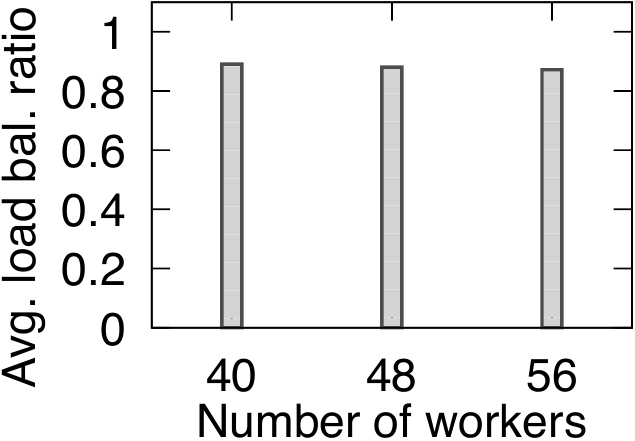}
  \caption{Texas data.}
  \label{fig:flink-tx}
     \end{subfigure}
     \vspace{-0.12in}
        \caption{Mitigation by \frmname on Flink ($W_1$ on tweets)}
        \label{fig:flink}
    \vspace{-0.15in}
\end{figure}

We implemented \frmname on Apache Flink and executed $W_1$ on $40$, $48$, and $56$ cores. A worker was classified as skewed if its {\em busyTimeMsPerSecond} metric was greater than $80$\%. Figure~\ref{fig:flink} shows the average load balancing ratio for the workers processing the California and Texas tweets. The ratio was about $0.9$, which means that the skewed and helper workers had similar workloads throughout the execution. For the $48$-core case, the final counts of tuples processed by the skewed and helper workers for California were $13$M and $14$M, respectively. The final counts of tuples processed by the two workers for Texas on $48$ cores were $10$M each. The execution time decreased as a result of the mitigation. For example, for the $48$-core case, the execution time decreased from $407$ seconds to $320$ seconds. 

%% file: sec8.tex
\section{Conclusions}
\label{sec:conclusions}

In this paper we presented a framework called \frmname that adaptively handles partitioning skew in the exploratory data analysis setting. We presented different approaches for load transfer and analyzed their impact on the results shown to the user. We presented an analysis about the effect of the skew-detection threshold on mitigation and used it to adaptively adjust the threshold. We generalized \frmname to multiple operators and broader execution settings. We implemented \frmname on top of two big data engines and presented the results of an experimental evaluation.